\documentclass[12pt,draftcls,onecolumn]{IEEEtran}

\ifCLASSINFOpdf

\else

\fi

\usepackage{graphics} 
\usepackage{amsmath} 
\usepackage{amssymb} 
\usepackage{breqn} 
\usepackage{cite}
\usepackage{algorithm}
\usepackage{algpseudocode}
\usepackage{pifont}
\usepackage{float}
\usepackage{lipsum}
\usepackage{graphicx}
\usepackage{pgf}
\usepackage{bbm}
\usepackage{etoolbox}
\usepackage{mathrsfs}
\usepackage{enumerate}
\usepackage{color}
\usepackage[textsize=footnotesize,textwidth=1.5cm]{todonotes}	

\newtheorem{thm}{Theorem}
\newtheorem{assumption}{Assumption}
\newtheorem{lem}[thm]{Lemma}
\newtheorem{definition}{Definition}
\newtheorem{rem}{Remark}

\newcommand{\argmin}{\operatornamewithlimits{argmin}}

\usepackage{soul}

\newcommand{\rma}[1]{\textcolor{black}{#1}}
\newcommand{\rsa}[1]{\textcolor{black}{#1}}
\usepackage{soul}

\newtoggle{jou}
\togglefalse{jou}

\begin{document}

\title{ 
Collective Stochastic Discrete Choice Problems:\\ A Min-LQG Game Formulation}

\author{Rabih~Salhab, 
        Roland~P.~Malham\'e               
        and~Jerome~Le~Ny% <-this % stops a space
\thanks{This work was supported by NSERC under Grants 6820-2011 and 435905-13.
The authors are with the department of Electrical Engineering, 
Polytechnique Montreal and with GERAD, Montreal, QC H3T-1J4, 
Canada {\tt\small \{rabih.salhab, roland.malhame, jerome.le-ny\}@polymtl.ca}.
}}

\markboth{}%
{Rabih \MakeLowercase{\textit{et al.}}: Bare Demo of IEEEtran.cls for Journals}

\maketitle

\begin{abstract}
%\jln{[Review LaTex packages used, check warnings remove those unused. The
%bibliography format is weird (underlined references).]}
%\jln{[Aim to get people interested with your abstract. For example: do you
%want to put immediately when you start "within the framework of MFG theory"? 
%The result will be that people will pass if they don't know about MFG. 
%Other example: no need to talk about dynamic programming or the maximum principle
%here, unless you want the readers to know that there is something new and
%interesting about these tools that you introduce in the paper.]}

We consider a class of dynamic collective choice models with social interactions, 
whereby a large number of non-uniform agents have to individually settle on one 
of multiple discrete alternative choices, with the relevance of their would-be 
choices continuously impacted by noise and the unfolding group behavior. 
For example, a large collection of geographically dispersed micro robots 
might need to gather, in a self organized mode and with least energy expenditure, 
at various known potential sites of interest, where groups of sufficient size 
would be required to perform collective tasks, such as search and rescue. 
This class of problems is modeled here as a so-called Min-LQG game, 
i.e., a linear quadratic Gaussian dynamic and non-cooperative game, 
with an additional combinatorial aspect in that it includes a final 
choice-related minimization in its terminal cost. 
The presence of this minimization term is \textit{key to enforcing 
some specific  discrete choice} by each individual agent.
The theory of mean field games is invoked to generate a class of 
decentralized agent feedback control strategies which are then shown 
to converge to an exact Nash equilibrium of the game as the number 
of  players increases to infinity. 
A key building block in our approach is an explicit solution to the 
problem of computing the best response of a generic agent to some 
arbitrarily posited smooth  mean field trajectory. 
Ultimately, an agent is shown to face  a continuously revised 
discrete choice problem, where greedy choices dictated by current 
conditions must be constantly balanced against the risk of the future 
process noise upsetting the wisdom of such decisions.
Even though an agent's ultimately chosen alternative is random 
and dictated by its entire noise history and initial state, 
the limiting infinite population macroscopic behavior can still be predicted. 
It is shown that any Nash equilibrium of the game is defined by 
an a priori computable probability matrix characterizing the manner 
in which the agent population ultimately splits among the available alternatives.  
For the scalar binary choice case with uniform dynamics 
and cost functions, we propose a numerical scheme to compute this probability matrix  %population
and the associated decentralized Nash strategies. 
The results are illustrated through simulations. 
\end{abstract}


\IEEEpeerreviewmaketitle

% !TEX root =  main.tex

\section{Introduction} \label{Intro}

Discrete choice problems arise in situations where an individual makes a choice 
among a discrete set of alternatives, such as modes of transportation \cite{Kappelman2005}, 
entry and withdrawal from the labor market, or residential locations \cite{Bhat2004}. 
In some situations, the individuals' choices are considerably influenced by the 
social behavior. For example, in schools, teenagers' decisions to smoke
are affected by some personal factors, as well as by their peers' behavior 
\cite{nakajima2007measuring}.

%\subsection{Related Work on Discrete Choice Problems}

In this paper, we study a discrete choice problem for a large population
of agents in a dynamic setting, capturing in particular the influence of the population's
state and the efforts that must be exerted over time by an individual before reaching
a decision.
For example, as in \cite{vail2003multi,halasz2007dynamic, hsieh2008biologically}, 
a group of robots exploring an unknown terrain might need to move within a 
finite time horizon from their initial positions towards one of multiple sites of interest. 
While moving, each robot optimizes its efforts to remain close to the group 
and to arrive at the end of the time horizon in the vicinity of one of the predefined sites. 
The group may split, but the size of the subgroups should remain large enough 
to perform some collective tasks, such as search and rescue. %, etc. 
Our model could also be used to study a mechanistic representation of opinion
crystallization in elections \cite{Merill99_vote, markus1979dynamic}, 
where voters continuously update their opinions until forming a final
decision regarding who they should vote for.
Along the path to choose a candidate, changing one's decision requires an effort, 
but at the same time, deviations from the others' opinions involve a discomfort.

McFadden laid in \cite{mcfadden1973conditional} the theoretical foundations of static
discrete choice models, where an agent chooses among a finite set of alternatives the
one that maximizes its utility. 
This utility depends on some potentially observable attributes (such as the agent's 
financial status in the example of residential location choice) that dictate 
a deterministic trend in the choice, and other attributes idiosyncratic to that agent, 
which are not known by the economist or social planner carrying out the macroscopic analysis,
although they may influence the choice.
As a result, utility is defined as the sum of a known deterministic term plus a random term. 
Later, Rust \cite{rust1994structural} introduced  
a dynamic discrete choice model involving a Markov decision process for each agent. 
While peer pressure effects are absent in Rust's and McFadden's models, 
Brock and Durlauf \cite{Brock2001} discuss a discrete choice problem with 
\textit{social interactions} modeled as a \textit{static} noncooperative game,
where a large number of agents have to choose between two 
alternatives while being influenced by the average of the choices.
The authors analyze the model using an approach similar to that of 
a static Mean Field Game (MFG), and inspired by statistical mechanics.

The Brock-Durlauf model includes peer influence but is static, 
and in Rust's model, the agents are required to make a choice repeatedly 
at each discrete time period but under no social influence. 
In our model, the agents are instead continuously reassessing the adequacy 
of their would-be choices and current actions along their random state-space path, 
up until the end of the control horizon, at which point their ultimate choice 
of alternative becomes fully  crystallized. 
Thus, our formulation helps in modeling situations where alternative choices 
are identified as potential destination points in a suitable state space 
(e.g., physical space in the robotics example, or opinion space in the election example), 
and implementation of a given choice involves movement towards a final destination state, 
requiring control effort and constrained by specific dynamics.

Recently, we introduced a related dynamic collective choice model with social interactions 
in \cite{Rab2014, Rab2015, DBLP:journals/corr/SalhabMN15}.
In these articles, we study using the MFG methodology a dynamic game involving 
a large number of players choosing between multiple destination points. 
The agents' dynamics are deterministic with random initial conditions. 
We show that multiple approximate (epsilon) Nash equilibria may exist, 
each characterized by a vector describing the way the population
splits between the alternatives. The strategies developed in these papers
are \textit{open loop} decentralized policies, in the sense that to make its choice
of trajectory and destination, an agent needs to know only its initial 
state and the initial distribution of the population. 
In particular, in this formulation each agent can commit from the start 
to its final choice, which implies that the model with deterministic dynamics 
is not sufficiently rich to fully capture opinion change phenomena.
%In this paper, however, 
In contrast, we consider here the fully stochastic case where a large number of players 
moving according to a set of controlled diffusion processes should lie at time $T$ 
in the vicinity of one of multiple predefined destination points.  
Along their path, they are influenced by the average state of the population and 
try to develop as little effort as possible to approach one of these destinations. 
By introducing noise in the dynamics we can model for example the unexpected 
events that influence a voter's opinion during electoral campaigns
or the random forces that perturb a robot's trajectory while choosing a site to visit.
The presence of this noise requires the agents to use bona fide feedback strategies
and prevents them from committing to a choice before the final time.
%However, one can still anticipate the manner in which a population growing to infinity, asymptotically splits among the set of alternatives.
However, one can still anticipate asymptotically the manner in which an infinite population 
splits among the set of alternatives.

\subsection{Mean Field Game Methodology and Contributions of This Paper}

The MFG methodology that we again follow in this paper  
was introduced in a series of papers by Huang {\it{et al.}} 
\cite{Huang03_wirelessPower, Huang07_Large, huang2006large}, 
and independently by Lions and Lasry \cite{lasry2006jeux, lasry2006jeux2, Lasry07_MFG}. 
It is a powerful technique to analyze dynamic games involving a large number of agents anonymously
interacting through their \textit{mean field}, i.e., the empirical distributions of the agents' states. Analysis starts by considering the limiting case of a continuum of agents. 
For agents evolving according to diffusion processes, the equilibrium of the game 
can be shown to be  characterized by the solution of two coupled partial differential equations (PDE), 
a Hamilton-Jacobi-Bellman (HJB) equation propagating backwards and a Fokker-Planck (FP) equation 
propagating forwards. 
Indeed, the former characterizes the agents' best response to some posited candidate 
mean field trajectory, while the latter propagates the would-be resulting mean field 
when all agents implement the computed best responses. 
Consistency requires that sustainable mean field trajectories, if they exist, 
be replicated in the process. %Thus, the requirement that limiting  equilibria 
Limiting equilibria are thus required to 
satisfy a system of fixed point equations, herein given by the coupled HJB-FP equations. 
The corresponding best responses, when applied to the finite population, constitute 
approximate Nash equilibria ($\epsilon$-Nash equilibria) \cite{huang2006large, Huang07_Large},
in the following sense.
\begin{definition}
Consider $N$ agents, a set of strategy profiles $S=S_1 \times \dots \times S_N$, and 
for each agent $i$, a cost function $J_{i}(u_1,\dots,u_N)$, for $(u_1,\dots,u_N) \in S$. 
A strategy profile $(u_1^*,\dots,u_N^*)\in S$ is called an $\epsilon$-Nash equilibrium 
with respect to the costs $J_{i}$ if there exists an $\epsilon>0$ such that, 
for any fixed $1\leq i \leq N$, for all $u_{i}\in S_i$, we have
$
J_i(u_i,u^*_{-i})\geq J_i(u_i^*,u_{-i}^*)-\epsilon.
$
\end{definition}

The main contributions and structure of the paper are described next.
In Section \ref{section_model}, we present the mathematical formulation 
of our multi-agent dynamic discrete choice problem. 
Then in Section \ref{section-Min-LQG}, we postulate a known trajectory 
for the agents population mean state, and compute the best response 
strategy of a generic agent. This involves solving a modified version of 
the standard Linear Quadratic Gaussian (LQG) optimal tracking problem, 
which we call the \emph{Min-LQG} optimal tracking problem, where the 
final cost is replaced by a minimum of $l$ distances to capture the 
discrete choice between $l$ alternatives.
We give an explicit expression for the generic agent's best response 
(Min-LQG optimal control law), and show that at each instant, 
it can be interpreted as a static discrete choice model \cite{mcfadden1973conditional}, 
where the cost for the agent of choosing one of the alternatives 
includes an additional term that increases with the risk of being 
driven to the other alternatives by the process noise.

In Section \ref{section-MFG-equ} we introduce the mean field equations, 
which allow us to compute the admissible population distribution 
trajectories. By exploiting the relationship between dynamic programming 
and the stochastic maximum principle, we establish a one-to-one map between 
the solutions of the mean field equations and the fixed points of a finite dimensional map. 
The latter characterizes the way the population splits between the alternatives. 
We show the existence of a fixed point for this map and propose in the binary choice case 
a numerical scheme to compute it. 
Our model lies somewhere between the MFG LQG models studied 
in \cite{Huang03_wirelessPower, Huang07_Large} and the fully nonlinear MFGs 
considered in \cite{huang2006large, lasry2006jeux, lasry2006jeux2, Lasry07_MFG}. 
On the one hand, in the LQG case the optimal strategies are linear feedback policies, 
whereas in our case they are nonlinear. On the other hand, an explicit solution in 
the general nonlinear case is not possible, while we provide in this paper an explicit 
form of the optimal strategies. Moreover, the existence and uniqueness of solutions 
of the mean field equations are shown in the general case under some strong regularity and 
boundedness conditions on the dynamics and cost coefficients \cite{huang2006large, lasry2006jeux, lasry2006jeux2, Lasry07_MFG}. In this paper, however, the quadratic running cost 
and linear dynamics make it possible to relax these assumptions. 

%
%We give in 
Finally, Section \ref{sim} discusses some numerical simulations illustrating the results, and
%results, while
Section \ref{conclusion} presents our conclusions.
Overall, the paper develops a set of decentralized closed-loop strategies %, which constitute 
forming an approximate Nash equilibrium for our multi-agent dynamic discrete choice problem. 
This equilibrium converges to an exact Nash equilibrium as the size of the population 
increases to infinity. To compute its optimal control at each time, an agent only needs 
to know its current individual state and the \emph{initial} distribution of the population, 
which could be learned by a consensus-like algorithm for example \cite{fagnani2008randomized}.
The rationality assumption in MFGs allows us to drastically reduce the amount of 
coordination needed between the players, by predicting the evolution of the population
distribution. In practice, with a finite number of players, errors in this prediction 
build up over time, but could be corrected by periodic communication between agents to 
re-estimate their current distribution.

\subsection{Notation}

Given an Euclidean space $X$, $B(c,r)$ denotes the ball of radius $r$ 
centered at $c$, $x_k$ the k-th element of $x \in X$, and $x_{-i}$ the vector
$(x_1,\dots,x_{i-1},x_{i+1},\dots,x_N)$. $\otimes$ denotes the Kronecker product. 
The $n \times n$ identity matrix is $I_n$, $M'$ denotes
the transpose of a matrix $M$, $\text{Tr}(M)$ its trace and  $|M|$ its determinant. 
The notations $M \succ 0$ and $M \succeq 0$ stand respectively for $M$ positive definite 
and positive semi-definite. 
Given an $n \times n$ matrix $Q \succeq 0$ and $x \in \mathbb{R}^n$, $\sqrt{\frac{1}{2}x'Qx}$ 
is denoted by $\|x\|_Q$. The minimum and maximum eigenvalues of a square matrix $Q$ are written 
$\lambda_{\min}(Q)$ and $\lambda_{\max}(Q)$.
Let $X$ and $Y$ be two subsets of Euclidean spaces. The set of functions from $X$ to $Y$ is
denoted by $Y^X$. $\partial A$, $\text{Int}(A)$ and 
$\bar{A}$ denote respectively the boundary, interior and closure of a subset $A \subset X$.
$C(X)$ refers to the set of $\mathbb{R}^n$-valued continuous functions on $X$ 
with the standard supremum norm $\|.\|_\infty$, and $C^{i,j}(X \times Y)$ to the set of real-valued continuous functions $f(x,y)$ on $X\times Y$ 
such that the first $i$ partial derivatives with respect to the 
$x$ variable and the first $j$ partial derivatives with respect to the $y$ variable are continuous. 
%\jlnComment{try to remove this sentence here and integrate with paragraph below}
%\jlnComment{give ref., perhaps Karatzas and Shreve?} 
The normal distribution of mean $\mu$ and covariance matrix $\Sigma$ is denoted 
by $\mathcal{N}(\mu,\Sigma)$. 

\section{Mathematical Model} \label{section_model}

We present in this section the dynamic collective choice model with social interactions.
We consider a dynamic noncooperative game involving a large number $N$ of players
with the following stochastic dynamics:
\begin{align} \label{agents_dyn}
dx_i(t)=\left(A_i x_i(t) + B_i u_i(t)\right)dt+\sigma_i
dw_i(t),
\end{align}
for $1\leq i \leq N$,
where $A_i \in \mathbb{R}^{n \times n}$, $B_i \in \mathbb{R}^{n \times m}$, 
$\sigma_i \in \mathbb{R}^{n \times n}$, and
$\{w_i,1 \leq i \leq N\}$ are $N$ independent Brownian motions in $\mathbb{R}^n$ on a probability space $(\Omega,\mathcal{F},\mathbb{P},\{\mathcal{F}_t\}_{t \in [0,T]})$, where $\{\mathcal{F}_t\}_{t \in [0,T]}$ is the Brownian filtration
generated by $\{w_i,1 \leq i \leq N\}$
\cite[Section 2.7]{karatzas2012brownian}.
In the remaining of the paper, $\mathbb{P}(A)$ denotes the probability 
of an event $A$, $\mathbb{E}(X)$ the expectation of a random variable $X$ and $\mathcal{M}([0,T],\mathbb{R}^n)$ the set of progressively measurable $\mathbb{R}^n$-valued functions with respect to 
the filtration $\{\mathcal{F}_t\}_{t \in [0,T]}$.
We assume that the matrices $\sigma_i$, 
$1 \leq i \leq N$, are invertible and
that the initial conditions $\{x_i(0),1 \leq i \leq N\}$ are 
independent and identically distributed (i.i.d.) and also independent
of $\{w_i,1 \leq i \leq N\}$. Moreover, we assume that
$\mathbb{E} \|x_i(0)\|^2 < \infty$. 
The vector $x_i(t) \in \mathbb{R}^n$ is the state of player $i$ at time $t$ and 
$u_i(t) \in \mathbb{R}^m$ its control input. 
Let $p_j$, $1 \leq j \leq l$, be $l$ points in $\mathbb R^n$.
Each player is associated with the following individual cost functional:
%\rmaComment{I think you are missing a 1/2 factor in your definition of the cost. 
%J: there is a 1/2 in his def. of $\|\cdot\|_M$}
\iftoggle{jou}{
\begin{align} \label{agents_cost} 
J_i\left(x_i(0),u_i,
u_{-i}\right)&=\mathbb{E} \bigg [
\int_0^T \left\{ \|x_i-\bar{x}\|^2_{Q_i} + \|u_i\|^2_{R_i} \right\} dt \nonumber\\ 
&+\min\limits_{1 \leq j \leq l} \|x_i(T)-p_j\|^2_{M_i} \; \Big | \; x_i(0)\bigg], 
\end{align}}{
\begin{equation} \label{agents_cost} 
J_i\left(x_i(0),u_i,
u_{-i}\right)=\mathbb{E} \left [
\int_0^T \left\{ \|x_i-\bar{x}\|^2_{Q_i} + \|u_i\|^2_{R_i} \right\} dt + 
\min\limits_{1 \leq j \leq l} \|x_i(T)-p_j\|^2_{M_i} \; \Big | \; x_i(0)\right], 
\end{equation}}
where $T>0$, $Q_i \succeq 0$, $R_i \succ 0$ and $M_i \succ 0$.
Along the path, the running cost encourages the players to remain grouped around  
the average of the population $\bar{x}(t):=\sum_{i=1}^N x_i(t)/N$, 
which captures the social effect, and to develop as little effort as possible. 
The form of the final cost captures the discrete choice aspect, if we assume 
$M_i$ large.
That is,
a player $i$ at time $T$ should be 
close to one of the destination 
points 
$p_j \in \mathbb{R}^n$, otherwise it is strongly penalized. 
%, in the sense of Lemma \ref{contr} below (see the discussion after
%Lemma \ref{contr}).
Hence, the overall individual cost captures the problem faced by each agent 
of deciding between $l$ alternatives, while trying to remain close to the mean 
population trajectory.

The players interact through the average $\bar{x}$ of the population. 
Even though the running cost is linear and quadratic, the minimum term 
in the final cost changes drastically the analysis of the game with 
respect to the standard LQG MFG problems \cite{Huang07_Large, Huang03_wirelessPower}. 
In fact, as shown later in (\ref{optimal-crt-exp}), the optimal strategies 
are nonlinear feedback policies, whereas they are linear in the standard LQG MFG problems.
Moreover, to anticipate the mean trajectory $\bar{x}$, the players 
need in the LQG case to know only the initial mean of the population. 
In our case, they need to know the full initial distribution in order to 
compute a solution of the nonlinear mean field equations defined later 
in (\ref{finite-operator}).

Let us denote the individual parameters $\theta_i:=(A_i,B_i,\sigma_i,Q_i,R_i,M_i)$. 
We assume that there are $k$ types of agents, that is, $\theta_i$ takes values in 
a finite set  $\{\Theta_1,\dots,\Theta_k\}$, which does not depend on the size 
of the population $N$. As $N$ tends to infinity,
it is convenient to represent the limiting sequence
of $(\theta_i)_{i=1,\dots,N}$ by realizations of a random
vector $\theta$, which takes values in the same finite set $\{\Theta_1,\dots,\Theta_k\}$.  
Let us denote the empirical measure of the sequence $\theta_i$ as    
$P^N_\theta(\Theta_s) = 1/N 
\sum_{i=1}^N 1_{\{\theta_i = \Theta_s
\}}$ for 
$s=1,\dots,k$. 
We assume that $P^N_\theta$ has a weak limit $P_\theta=(\alpha_1,\dots,\alpha_k)$ as $N \to \infty$ 
% \rmaComment{Would alphas work instead of n's which suggest integers?}.
For further discussions about this assumption, one can refer to \cite{huang2012social}.

We define the set of admissible control laws for each agent as follows
\begin{equation}
\begin{split}
\mathcal{U}=\left \{ u \in \mathcal{M}([0,T],\mathbb{R}^m) \; \Big | \; \mathbb{E} \int_0^T \|u(s)\|^2 ds < \infty, 
 \right\}.
\end{split}
\end{equation}
We define the set of admissible Markov policies
\iftoggle{jou}{
\begin{align} \label{set_l}
\begin{split}
&\mathcal{L}=\Big \{ u \in 
(\mathbb{R}^m)^{[0,T]\times \mathbb{R}^n} 
| \exists L_1>0, \forall (t,x) \in [0,T]
\times \mathbb{R}^n, \\
&\|u(t,x)\| \leq L_1(1+\|x\|),
\text{ and } \forall r >0, T' \in (0,T),
\exists L_2>0, \\
&\forall \|(x,y)\| \leq r,
 t\in [0,T'], \|u(t,x)-u(t,y)\| \leq L_2
\|x-y\|\Big 
 \}.
\end{split}
\end{align}}
{\begin{equation} \label{set_l}
\begin{split}
&\mathcal{L}=\Big \{ u \in 
(\mathbb{R}^m)^{[0,T]\times \mathbb{R}^n} 
| \exists L_1>0, \forall (t,x) \in [0,T]
\times \mathbb{R}^n, \|u(t,x)\| \leq L_1(1+\|x\|),
\text{ and } \\
&\forall r >0, T' \in (0,T),
\exists L_2>0, \forall \|(x,y)\| \leq r,
 t\in [0,T'], \|u(t,x)-u(t,y)\| \leq L_2
\|x-y\|\Big 
 \}.
\end{split}
\end{equation}}
If $u \in \mathcal{L}$, then the
stochastic differential equation (SDE) 
(\ref{agents_dyn}), with $u_i$ equal 
to $u(t,x_i)$, has a unique 
strong solution \cite[Section 5.2]{karatzas2012brownian}. 
As shown below in Theorem \ref{epsilon-Nash}, the developed strategies form
$\epsilon-$Nash equilibria with respect to the space of admissible actions $\mathcal{U}$. 
Moreover, these strategies can be expressed as Markov policies (feedback policies),
see \eqref{optimal-crt-exp}. 
%\pagebreak

\section{The ``Min-LQG" Optimal Tracking Problem\\ and the Generic Agent's Best Response} \label{section-Min-LQG}

Following the MFG approach, we start by assuming a continuum of
agents for which one can ascribe a \emph{deterministic} macroscopic 
behavior $\bar{x}$ (posited mean population state), which is supposed known in this section.
The problem of determining $\bar x$ is treated in Section \ref{section-MFG-equ}.
In order to compute its best response to $\bar{x}$, a generic agent with parameters $\theta = 
(A,B,\sigma,Q,R,M) \in \{\Theta_1,\dots,\Theta_k\}$ solves the following 
optimal control problem, which we call the ``Min-LQG" optimal tracking problem:
\iftoggle{jou}{
\begin{equation} \label{min-LQG}
\begin{split}
\inf\limits_{u\in \mathcal{U}} 
J\left(x(0),u,\bar{x}\right)&=
\inf\limits_{u\in \mathcal{U}}\mathbb{E} \Big [
\int_0^T \left\{ \|x-\bar{x}\|^2_Q
+ \| u\|_R^2 \right\} dt \\
&+ 
\min\limits_{j=1,\dots,l} \|x(T)-p_j\|_M^2 \Big | x(0)\Big]\\
\text{s.t. }
dx(t)&=\left(A x(t) + Bu(t)\right)dt+\sigma
dw(t),
\end{split}
\end{equation}}{
\begin{equation} \label{min-LQG}
\begin{split}
&\inf\limits_{u\in \mathcal{U}} 
J\left(x(0),u,\bar{x}\right)=
\inf\limits_{u\in \mathcal{U}}\mathbb{E} \left [
\int_0^T \left\{ \|x-\bar{x}\|^2_Q
+ \| u\|_R^2 \right\} dt + 
\min\limits_{j=1,\dots,l} \|x(T)-p_j\|_M^2 \Big | x(0)\right]\\
&\text{s.t. }
dx(t)=\left(A x(t) + Bu(t)\right)dt+\sigma
dw(t),
\end{split}
\end{equation}}
where $w$ is a Brownian motion in $\mathbb{R}^n$ on the probability space
$(\Omega,\mathcal{F},\mathbb{P})$ and $x(0)$ is a random vector independent of $w$ and
distributed according to the known distribution of the agents' initial states. \rma{ Herein,}\rsa{ $\frac{\partial h(t,x)}{\partial x}$ and 
$\frac{\partial^2 h(t,x)}{\partial x^2}$ will denote respectively the gradient and Hessian matrix of the real function $h$ with respect to $x \in \mathbb{R}^n$.}
The optimal cost-to-go function of (\ref{min-LQG}) satisfies the following
HJB equation \cite{fleming2006controlled}
\iftoggle{jou}{
\begin{equation} \label{general_HJB}
\begin{split}
-\frac{\partial V}{\partial t} &= 
x'A'\frac{\partial V}{\partial x}
-\frac{1}{2}\left( 
\frac{\partial V}{\partial x} \right)'BR^{-1}B' \frac{\partial V}{\partial x} \\
&+
\frac{1}{2} \text{Tr} \left( \sigma' 
\frac{\partial^2 V}{\partial x^2} \sigma \right)
+\|x-\bar{x}\|^2_Q\\
V(T,x)&= \min\limits_{1 \leq j \leq l} 
\|x-p_j\|^2_M, \,\, \forall x \in \mathbb{R}^n. 
\end{split}
\end{equation}}{
\begin{equation} \label{general_HJB}
\begin{split}
-\frac{\partial V}{\partial t} &= 
x'A'\frac{\partial V}{\partial x}
-\frac{1}{2}\left( 
\frac{\partial V}{\partial x} \right)'BR^{-1}B' \frac{\partial V}{\partial x} +
\frac{1}{2} \text{Tr} \left( \sigma' 
\frac{\partial^2 V}{\partial x^2} \sigma \right)
+\|x-\bar{x}\|^2_Q\\
V(T,x)&= \min\limits_{1 \leq j \leq l} 
\|x-p_j\|^2_M, \,\, \forall x \in \mathbb{R}^n. 
\end{split}
\end{equation}} 
In the following, we linearize equation \eqref{general_HJB} under appropriate conditions,
using a generalized Hopf-Cole transformation \cite[Section $4.4$]{evans1998partial}.
Moreover, we derive an explicit formula for the 
solution of \eqref{general_HJB} and the 
min-LQG optimal control law. 
The results of this section are proved in Appendix \ref{ProofA}.

The following notation is used in the remaining of the paper. %entities are
We define $x^{(j)}$, $u^{(j)}$ and $V_j$ to be the optimal state trajectory, 
optimal control law and optimal cost-to-go of the LQG tracking problem that the
generic agent must solve when $p_j$ is the only available alternative, that is,
\eqref{min-LQG} with $p_k=p_j$, for all 
$k \in \{1,\dots,l\}$. 
Recall that \cite[Chapter 6]{yong1999stochastic} 
\begin{align}
V_j(t,x)&=\frac{1}{2}x'\Pi(t)x+x'\beta_j(t)+\delta_j(t) \label{aux_cost}\\
u^{(j)}(t,x)&=-R^{-1}B'\left (\Pi(t)x+\beta_j(t)\right) \label{aux_ctr}\\
dx^{(j)}(t)&=\left(A x^{(j)}(t) + Bu^{(j)}
\left(t,x^{(j)}(t)\right)\right)dt+\sigma dw^{(j)}(t), \label{aux_st}
\end{align}
where $\Pi$, $\beta_j$ and $\delta_j$ are the unique solutions of
\begin{align} 
\frac{d}{dt}\Pi(t) &= \Pi(t)BR^{-1}B'\Pi(t)-A'\Pi(t)-\Pi(t)A -Q,\nonumber\\
\frac{d}{dt}\beta_j(t)&=-\left(A-BR^{-1}B'\Pi(t)\right)' 
\beta_j(t)+Q\bar{x}(t), \label{pi-beta-delta}\\
\frac{d}{dt}\delta_j(t)&=\frac{1}{2}
\beta_j(t)'BR^{-1}B'\beta_j(t)-\frac{1}{2}\text{Tr}(\sigma'\Pi(t)\sigma)
- \|\bar{x}(t)\|_Q^2,\nonumber
\end{align}
with $\Pi(T)=M,\beta_j(T)=-Mp_j$ and $\delta_j(T) =\|p_j\|^2_M $.
\begin{rem}
% \rmaComment{The above boundary conditions need fixing. Last term divided by 2?}
The final cost in \eqref{min-LQG} is non-smooth. Hence, the 
corresponding HJB equation 
\eqref{general_HJB} and its transformed parabolic equation \eqref{parabolic} below 
have non-smooth terminal conditions. %boundary conditions. 
However, as shown later in Lemma \ref{lem-cont}, these partial differential equations (PDEs) 
smooth out their solutions, i.e., the only non-smoothness %is in the boundary condition. 
occurs at the terminal time.
Hence, all the PDE solutions in the remaining sections are to be
understood in the strong sense.
\end{rem}

In the following, we denote by $\mathcal{W}_j$, for $j \in \{1,\dots,l\}$, the Voronoi cell 
associated with $p_j$, that is, 
$\mathcal{W}_j=\{x \in \mathbb{R}^n \text{ such that } 
\|x-p_j\|_M \leq \|x-p_k\|_M, \text{ for all } 1 \leq k \leq l \}$.
We define for all $j \in \{1,\dots,l\}$ the following notation for the conditional probability
of an agent following the control law $u^{(j)}$ to be in the Voronoi cell $j$ at time $T$ 
if its state at time $t$ is $x$
\begin{equation} \label{g1_g2}
g_j(t,x) \triangleq \mathbb{P}\left (x^{(j)}(T) \in \mathcal{W}_j \Big| x^{(j)}(t)=x\right),
\end{equation}
where $x^{(j)}$ is defined in \eqref{aux_st}.
To linearize and solve the HJB equation (\ref{general_HJB}) using the 
Hopf-Cole transformation (see Appendix \ref{ProofA}), we make the following assumption. %need
\begin{assumption} \label{assumption_linear}
We assume that there exists a scalar $\eta > 0$ such that 
$B R^{-1}B'=\eta\sigma \sigma'$.
% $\eta \in \mathbb{R}$
\end{assumption}
\begin{rem}
Note the following:

\begin{enumerate}[i.] 
\item Assumption \ref{assumption_linear}
always holds in the scalar case 
($n=m=1$).
\item If $\eta$ exists, then it is  strictly positive.
% \item If $\eta$ exists, then the dimension of control space is greater 
% \jln{or equal to} that of the state space ($m \geq n$). In fact, we assumed that 
% $\sigma$ is invertible. Hence, if $\eta$
% exists, then $BR^{-1}B'$ is invertible, and thus $\text{ker } B' = \{0\}$. 
\item If $\eta$ exists, since we assumed that $\sigma$ is invertible,
so is $BR^{-1}B'$, and thus $\text{ker } B' = \{0\}$.
In particular, the dimension of control space is greater 
or equal to that of the state space ($m \geq n$). 
%In fact, we assumed that $\sigma$ is invertible. Hence, if $\eta$
%exists, then $BR^{-1}B'$ is invertible, and thus $\text{ker} B' = \{0\}$. 
Then we must choose $R = \frac{1}{\eta} B'(\sigma \sigma')^{-1} B$, for
some $\eta > 0$.
\item Assumption \ref{assumption_linear} is satisfied in particular if 
$B = R = \sigma = I_n$, a situation that has been studied previously in the context 
of other mean-field games (with $A = 0$) using the Hopf-Cole transformation, 
see \cite[Chapter 2]{Gomes:Book16:regulatiryMFG} and the references therein.
\end{enumerate}
\end{rem}
We now state the main result of this section. 
%\jln{which is proved in Appendix \ref{appendix: pf best response}.} %\ref{Proof2}.}
\begin{thm} \label{thm-hjb}
Under Assumption \ref{assumption_linear}, the HJB equation (\ref{general_HJB})
has a unique solution $(t,x) \mapsto V(t,x)$ in 
$C^{1,2}([0,T)\times \mathbb{R}^n) \cap C([0,T] \times \mathbb{R}^n)$, defined as
\iftoggle{jou}{
\begin{equation} \label{sol_hjb}
\begin{split}
V(t,x)&= - \frac{1}{\eta}
\log \left( \sum_{j=1}^l \exp \left( -\eta V_j(t,x) \right) 
g_j(t,x)
\right),\\
&\forall (t,x) \in [0,T)\times
\mathbb{R}^n\\
V(T,x)&= 
\min\limits_{j=1,\dots,l} \|x-p_j\|_M^2, \,\,
\forall x \in \mathbb{R}^n,
\end{split}
\end{equation}}{
\begin{equation} \label{sol_hjb}
\begin{split}
V(t,x)&= - \frac{1}{\eta}
\log \left( \sum_{j=1}^l \exp \left( -\eta V_j(t,x) \right) 
g_j(t,x)
\right),
\forall (t,x) \in [0,T)\times
\mathbb{R}^n\\
V(T,x)&= 
\min\limits_{j=1,\dots,l} \|x-p_j\|_M^2, \,\,
\forall x \in \mathbb{R}^n,
\end{split}
\end{equation}
}
where $V_j$ and $x^{(j)}$ are 
defined in \eqref{aux_cost} and \eqref{aux_st}.
\end{thm}

Having solved the HJB equation related to 
the Min-LQG optimal control problem 
(\ref{min-LQG}), we now prove the 
existence of a unique optimal control 
law.
We define the following function:
\begin{equation} \label{optimal_ctr}
\begin{split}
u_*(t,x)&=-R^{-1}B'\frac{\partial V}{\partial x}, \;\;
\;\;\;\; t \in [0,T)
\\
%&=
%\frac{1}{\eta}R^{-1}B'
%\frac{\sum_{j=1}^l
%\exp \left( -\eta V_j(t,x) \right)
%}{
%\sum_{j=1}^l\exp \left( -
%\eta V_j(t,x) \right)
%g_j(t,x)} \left(-\eta\left(\Pi x + \beta_j\right)g_j(t,x)+
%\frac{\partial g_j}{\partial x}\right)\\
u_*(T,x)&=0.
\end{split}
\end{equation}
\begin{thm} \label{thm-opt}
The following statements hold:
\begin{enumerate}[i.]
\item The function $u_*$ defined in (\ref{optimal_ctr}) has on 
$[0,T)\times \mathbb{R}^n$ the following form: 
\begin{equation} \label{optimal-crt-exp}
u_*(t,x)=\sum_{j=1}^l\frac{\exp\left(- 
\eta V_j(t,x) \right) g_j(t,x)}{\sum_{k=1}^l\exp\left(- 
\eta V_k(t,x) \right)
g_k(t,x)}u^{(j)}(t,x),
\end{equation}
with $V_j, u^{(j)}$ defined in \eqref{aux_cost} and \eqref{aux_ctr}.
\item $u_*$ is an admissible Markov
policy.
\item $u_*\left(t,x_*(t,w)\right)$ is the unique 
optimal control law of (\ref{min-LQG}),
where $x_*(t,w)$ is the unique strong
solution of the SDE in (\ref{min-LQG})  
with $u$ equal to $u_*(t,x)$.
\end{enumerate}
\end{thm}

In the degenerate case with $\sigma=0$, it is shown in \cite{DBLP:journals/corr/SalhabMN15,Rab2014,Rab2015} that the optimal strategy 
of an agent $i$ in the Min-LQG problem is equal to $u^{(j)}$ 
(the optimal strategy in the presence of only one alternative $p_j$) if the 
Linear Quadratic Regulator (LQR) control problem associated with $p_j$ is 
the least costly starting from $x_i(0)$. 
Therefore, a generic agent commits from the start to its final choice
based on its initial state. When $\sigma\neq 0$, 
the generic agent can no longer be ``decisive". Its
optimal control law (\ref{optimal-crt-exp}) is a convex 
combination of the optimal policies $u^{(j)}$, $j=1,\dots,l$. The weights of
$u^{(j)}$ form a spatio-temporal Gibbs distribution \cite{liggett2012interacting}, 
which puts more mass on the less costly and risky destinations. 
A destination $p_j$ is considered riskier in  state $x$ at time $t$
if the Brownian motion has a higher chance of driving the state of an
agent closer to a destination different from $p_j$ at time $T$, when this
agent implements $u^{(j)}$ from $(x,t)$ onwards.

We claim in Section \ref{section_model} that 
the final cost forces the agents to be close 
to one of the destination points at time $T$. In the following Lemma, we show in fact that in the scalar case, the 
probability that an agent is close to one of the desitnation points increases with the final cost's coefficient $M$. The result is proved for paths 
$\bar{x}$ that 
are uniformly bounded with respect to $M$, a property that is shown to hold later in 
Theorem \ref{lemma:existance11} for the fixed point paths $\bar{x}$.  
\begin{lem} \label{contr}
Suppose that the paths $\bar{x}$ are uniformly bounded with respect to $M$
for the norm $\Big(\int_0^T \|.\|^2\mathrm{dt}\Big)^{\frac{1}{2}}$. Then, 
for any $\epsilon>0$, 
\[
\mathbb{P} \left(|x_*(T)-p_j|> \epsilon, 
\forall 1 \leq j \leq l\right)
= o\left(\frac{1}{M}+ \frac{\sigma^2}{2}\frac{\log M}{M} \right).\]
\end{lem}  

In the rest of this section, we discuss the relation between our solution to the Min-LQG optimal 
control problem in the scalar binary choice case and the solution of static discrete choice models. 
We start by recalling some facts about the static models. In the standard binary 
discrete choice models, a generic person chooses between two alternatives
$1$ and $2$. The cost paid by this person when choosing an alternative 
$j$ is defined by
$
v_{j}=k(j)+\nu,
$
%\jlnComment{changed $\epsilon$ to 
%$\nu$, too many $\epsilon$'s}
where $k(j)$ is a deterministic function that depends on personal publicly observable
attributes and on alternative $j$, while $\nu$ is a random variable accounting for personal 
idiosyncrasies unobservable by the social planner. 
When $\nu$ is distributed according to the extreme value 
distribution \cite{mcfadden1973conditional}, 
then the probability that a cost-minimizing generic person chooses 
an alternative $j$ is equal to
$
P_{j}=\frac{\exp(-k(j))}{\exp(-k(1))+\exp(-k(2))}.
$
Now, the Min-LQG optimal control law
(\ref{optimal-crt-exp}) can be 
written as follows:
\begin{align} \label{optimal-crt-exp1}
&u_*(t,x)=\frac{\exp\left(- 
\eta \tilde{V}_1(t,x) 
\right)}{\exp\left(- 
\eta \tilde{V}_1(t,x) 
\right)+\exp\left(- 
\eta \tilde{V}_2(t,x) 
\right)}u^{(1)}(t,x)\nonumber\\
&+\frac{\exp\left(- 
\eta \tilde{V}_2(t,x) \right)}{\exp\left(- 
\eta \tilde{V}_1(t,x) \right)+\exp\left(- 
\eta \tilde{V}_2(t,x) \right)}u^{(2)}(t,x),
\end{align}
where
\begin{align} \label{discrete-cost}
\tilde{V}_j(t,x)=V_j(t,x)-
\frac{1}{\eta}\log \left( 
g_j(t,x) \right), && j=1,2.
\end{align}
Here $V_j(t,x)$ is the expected cost paid by a generic agent 
if $p_j$ is the only available alternative, 
%. In this case, is the optimal policy. 
and $u^{(j)}(t,x)$ is the corresponding 
optimal policy.
In the presence of two alternatives, 
the optimal policy at time $t$ is given by (\ref{optimal-crt-exp1}), which
can be interpreted as a mixed strategy between two pure strategies 
$u^{(1)}(t,x)$ (picking alternative $p_1$) and $u^{(2)}(t,x)$ (picking 
alternative $p_2$). Within this framework, denoting by $-j$ the alternative 
other than $j$, a generic agent at time $t$ chooses the alternative $p_j$ with probability
\[
Pr_j=\frac{\exp\left(- 
\eta \tilde{V}_j(t,x) \right)}{\exp\left(- 
\eta \tilde{V}_j(t,x) \right)+\exp\left(- 
\eta \tilde{V}_{-j}(t,x) \right)}.
\]
Thus, the  
Min-LQG problem
can be viewed at each time $t \in [0,T]$ as a static discrete choice problem, 
where the cost of choosing alternative $p_j$ includes an additional term 
$-\frac{1}{\eta}\log \left( g_j(t,x) \right)$ which goes to zero when the 
probability $g_j(t,x)$ of landing inside of the Voronoi cell $\mathcal{W}_j$ 
associated with alternative $j$ at time $T$ approaches one. 
In effect, this additional term measures the expected cost due to the risk 
inherent to an  early choice, i.e., one based on a premature decision
at time $t<T$ in favor of one alternative, for example alternative $p_j$, 
thus applying all the way the pure strategy $u^{(j)}(t,x)$, while ignoring 
the possibility that by time $T$, the Brownian motion could drive the agent's 
state toward the $-j$ alternative.
\section{The Mean Field Equations and The Fixed Point Problem} \label{section-MFG-equ}

In Section \ref{section-Min-LQG}, we assumed the mean trajectory $\bar{x}$ known
%a given macroscopic behavior $\bar{x}$
and we computed the generic agent's best response to it, which is given by (\ref{optimal-crt-exp}). 
In the following, a superscript $s$ refers to an agent with parameters 
$\Theta_s \in \{\Theta_1,\dots,\Theta_k\}$. We write $u_*^s(t,x,\bar{x})$ 
instead of $u_*^s(t,x)$ to emphasize the dependence on $\bar{x}$. 
We now seek a sustainable macroscopic behavior $\bar{x}$, in the sense 
that it is replicated by the mean of the generic agent's state under 
its best response to it.
Thus, an admissible $\bar{x}$ satisfies the following Mean Field equations:
\begin{align} \label{gen_SDE} 
\bar{x}(t)&= \sum_{s=1}^k \alpha_s \bar{x}^s(t), \text{ with } \bar{x}^s=\mathbb{E}[x_*^s], \; 1 \leq s \leq k, \\
dx_*^s(t)&= (A^s x_*^s(t) + B^su_*^s \left(t,x_*^s(t),\bar{x})\right)dt+\sigma^s dw^s(t),\nonumber
\end{align}
%for $s=1,\dots,k$, 
where $x_*^s(0)$, $1\leq s \leq k$, are independent and identically distributed 
according to the initial distribution of the players, 
i.e., the distribution of $x_i(0)$, $\alpha_s$ are defined in Section \ref{section_model}
%$\bar{x}^s=\mathbb{E}x_*^s(t)$ 
and $\{w^1,\dots,w^k\}$ are $k$ independent Brownian motions, 
assumed independent of $\{x_*^1(0),\dots,x_*^k(0)\}$. 

Solving (\ref{gen_SDE}) directly is 
\rma{no} easy task.
Indeed, this is an
%Equation (\ref{gen_SDE}) is a 
$n \times k$ nonlinear Mckean-Vlasov 
equation \cite{huang2006large},
where the drift term depends on 
the joint probability law of the 
$k$ types of generic agents through 
the mean trajectory $\bar{x}$.  
\rsa{However, we derive in Lemma \ref{lem-equ} below 
an equivalent representation for a solution $\bar x$ of (\ref{gen_SDE}),
based on the solution of a \rma{deterministic optimal control} problem,
consisting of two linear forward-backward ODEs 
\rma{(see \eqref{MF-1}-\eqref{MF-2} below),}
with a nonlinear coupling in the 
boundary condition \rma{(see \eqref{MF-3}-\eqref{MF-4})} through 
what we call a ``Choice Distribution Matrix'' (CDM). 
A CDM is a $k \times l$ row stochastic matrix with 
its $(s,j)$ entry equal to the probability that the generic agent of type 
$s$ is at time $T$ closer \rma{(in the sense of the $M$-weighted $l_2$ norm)} 
to  $p_j$ than any of the other alternatives when it optimally 
responds to $\bar{x}$.
Equations \eqref{MF-1}-\eqref{MF-2} are respectively the \rma{expectations}
of the generic agents' state \rma{(see \eqref{agents_dyn})} and 
co-state equations \rma{ (see \eqref{adjoint_process}), together with that of their boundary conditions (see \eqref{limit_proof}),  when the input in \eqref{agents_dyn} associated with the agents' best response is expressed in terms of the co-state}. 
The state and co-state equations follow from the stochastic maximum principle
of the Min-LQG problem, which we derive in Lemma \ref{stochastic_max} in Appendix \ref{ProofB}. 
}
% \rsa{We derive at first a stochastic 
% maximum principle for the min-LQG problem
% (See Lemma \ref{stochastic_max} in Appendix \ref{ProofB} below). By aggregating the 
% corresponding state and co-state equations, we provide in Lemma
% \ref{lem-equ} a new representation of the 
% solutions of \eqref{gen_SDE}, which consists of  
% two coupled forward-backward ODEs
% \eqref{MF-1}-\eqref{MF-2}, with boundary 
% condition 
% that depends on what we call a ``Choice 
% distribution matrix (CDM)''. A CDM is 
% a $k \times l$ row stochastic matrix with 
% its $(s,j)$ entry equal to the 
% probability that the generic agent of type 
% $s$ is at time $T$ closer to  $p_j$ than any 
% of the other alternatives when it optimally 
% responds to $\bar{x}$.}
%
%Following Lemma \ref{lem-equ}, 
\rsa{The advantage of this new representation is that if one 
considers the CDM in the \rma{$\bar{q}(T)$} boundary
condition as a parameter (say any \rma{$k \times l$ }row 
stochastic matrix), 
then 
equations \eqref{MF-1}-\eqref{MF-2}  become totally linear.
As a result, they have
under Assumption \ref{assum_riccati} \rma{below}
an explicit solution (the term that 
multiplies $P_1$ in \eqref{eq: x-lambda def}) parametrized
by the stochastic matrix. This implies that
the solutions $\bar x$ of the 
mean field equations lies in the 
family of paths \eqref{eq: x-lambda def} parametrized by
the \rma{$k \times l$} row stochastic matrices. Conversely, in order to have a path $\bar x^\Lambda$ parametrized by \rma{some candidate} $\Lambda$ \rma{be}
a solution of \eqref{gen_SDE}, \rma{consistency requires that}
$\Lambda$  be equal to
the \rma{associated CDM as expressed in \eqref{MF-3}} when the generic agents 
optimally respond to $\bar x^\Lambda$. This is equivalent to 
\rma{requiring that} $\Lambda$ be a fixed point
of the finite dimensional map $F$ defined by \eqref{finit_ope}. 
Indeed, $F$ maps a row stochastic matrix $\Lambda$ to the CDM when 
the generic agents
optimally respond to $\bar x^\Lambda$. 
}

\rma{In effect, we establish that there is} a one to one map between 
the solutions $\bar{x}$ of (\ref{gen_SDE})
and the fixed points of the finite dimensional map $F$. \rma{Theorem \ref{lemma:existance11} below summarizes the related results:} 
\rsa{point (i) of the theorem  states the existence of the one-to-one map in question,} \rma{while point (ii)  states that  there is at least one  fixed point of $F$. Thus, a Nash equilibrium CDM exists}  (equivalently a solution of (\ref{gen_SDE})) \rma{characterizing}  the way 
an infinite population, \rma{with the same distribution of heterogeneous parameters as the original large but finite population,} would split between the destination points. 
The \rma{above} results  are proved in Appendix \ref{ProofB}. 

% In the following, we adopt the following notations: $\bar{X}=(\bar{x}^1,\dots,\bar{x}^k)$,
% $X=(x_*^1,\dots,x_*^k)$,
% $U=(u_*^1,\dots,u_*^k)$,
% $W=(w^1,\dots,w^k)$,
% $A=\text{diag}(A^1,\dots,A^k)$,
% $B=\text{diag}(B^1,\dots,B^k)$,
% $Q=\text{diag}(Q^1,\dots,Q^k)$,
% $R=\text{diag}(R^1,\dots,R^k)$,
% $M=\text{diag}(M^1,\dots,M^k)$,
% $\sigma=\text{diag}(\sigma^1,\dots,\sigma^k)$,
% $p=(p_1,\dots,p_l)$,
% and $L=I_{nk}- L_1\otimes I_n $,
% where $L_1$ is the $k \times k$  matrix with identical rows equal to
% $P_\theta=(n_1,\dots,n_k)$.
In the following, we adopt the following notations. 
%\jlnComment{I rewrote this 
%paragraph}
Let $\bar{X}=(\bar{x}^1,\dots,\bar{x}^k)$,
$X_*=(x_*^1,\dots,x_*^k)$,
$U_*=(u_*^1,\dots,u_*^k)$,
$W=(w^1,\dots,w^k)$, and $p=(p_1,\dots,p_l)$.
Let $A, B, Q, R, M$ and $\sigma$ be the block-diagonal
matrices $\text{diag}(A^1,\dots,A^k)$, $\text{diag}(B^1,\dots,B^k)$,
$\text{diag}(Q^1,\dots,Q^k)$, $\text{diag}(R^1,\dots,R^k)$, 
$\text{diag}(M^1,\dots,M^k)$ and $\text{diag}(\sigma^1,\dots,\sigma^k)$
respectively.
Define $L=I_{nk}- 1_k\otimes P_1$,
where $1_k$ is a column of $k$ ones and $P_1=P_\theta' \otimes 
I_n$ \rma{where $P_\theta$ is defined in Section \ref{section_model}}. 
\rsa{The following assumption guarantees 
the existence and uniqueness of a solution
for \eqref{MF-1}-\eqref{MF-2} whenever the 
CDM in the \rma{$\bar{q}(T)$ }boundary condition  is considered as a parameter.}
\begin{assumption} \label{assum_riccati}
We assume the existence of a solution on $[0,T]$ to the following 
(nonsymmetric) Riccati equation:
%\jlnComment{I find it a bit ambiguous the issue we are trying to %rule out
%here; existence, global existence, or unicity?}
\begin{equation} \label{aux_riccati}
\frac{d }{dt}\pi=-A'\pi-\pi A+\pi BR^{-1}B'\pi + QL, \quad \pi(T)=M.
\end{equation}
\end{assumption}
Note that if Assumption \ref{assum_riccati} is satisfied, the 
solution of \eqref{aux_riccati} is unique as a consequence of 
the smoothness of the right-hand side of \eqref{aux_riccati} 
with respect to $\pi$ \cite[Section 2.4, Lemma 1]{perko2013differential}.
For a uniform population, i.e., $k=1$, we have $L=0$ and hence 
Assumption \ref{assum_riccati} is always satisfied \cite[Section 2.3]{anderson2007optimal}. For more details 
about Assumption \ref{assum_riccati}, one can refer to \cite{freiling2002survey}.

\begin{lem} \label{lem-equ}
Under Assumption \ref{assum_riccati}, $\bar{x}$ satisfies 
(\ref{gen_SDE}) 
%\jlnComment{simply say $\bar{x}$ satisfies (17)? why fixed point?}
if and only if it satisfies the following equations 
%\rmaComment{Is (22) not a solution 
%of (19a)-(19c)? provided (18) has a 
%solution?}
%\jlnComment{Note: only coupling seems to be in (19b) - notation with diag. somewhat obfuscating}
\begin{subequations} 
\begin{align}
\frac{d}{dt}\bar{X}(t)&=A\bar{X}(t)-BR^{-1}B'
\bar{q}(t),
\label{MF-1}\\
\frac{d}{dt} \bar{q}(t)&=-A' \bar{q}(t)+QL\bar{X}(t),
\label{MF-2}\\
\bar{x}(t)&=P_1\bar{X}(t), \label{MFF}
\end{align}
\end{subequations}
with $\bar{X}(0)=\mathbb{E}X(0)$ and $\bar{q}(T)=M\left(\bar{X}(T)-
\Lambda \otimes I_n p\right)$,
where the CDM $\Lambda$ is defined
as follows:
\begin{subequations} 
\begin{align}
&\Lambda_{sj}=
\mathbb{P}(x_*^s(T) \in \mathcal{W}_j), \qquad 1 \leq s \leq k, \;\; 
1 \leq j \leq l
\label{MF-3} \\
&dX_*(t)= (A X_*(t) + BU_* \left(t,X_*(t),\bar{x})\right)dt+\sigma dW(t). \label{MF-4}
\end{align}
\end{subequations}
\end{lem}
We are now ready to state the main result of this paper, which asserts 
that there exists an admissible mean trajectory $\bar x$ satisfying (\ref{gen_SDE}). Moreover, 
it characterizes each admissible $\bar{x}$ by a matrix $\Lambda$, 
where $\Lambda_{sj}$ given in \eqref{MF-3} is the probability that an agent 
of type $s$ optimally tracking $\bar{x}$ is at time $T$ closer to $p_j$ than 
any of the other destination points. 
%Furthermore, 
This matrix is a fixed point of a well defined finite dimensional map.

\rsa{The following functions are 
used to compute the solution of \eqref{MF-1}-\eqref{MF-2}, where 
$\Lambda$ is considered as a parameter.}
Under Assumption \ref{assum_riccati}, we define $R_1$ and $R_2$ such that, 
%$\forall s \geq 0$,
for all $s \geq 0$,
\rsa{
\begin{equation} \label{eq:r1r2}
\begin{split}
\frac{d}{dt}R_1(t,s)&=\left(A-BR^{-1}B'\pi(t)\right)R_1(t,s),\\
\frac{d}{dt}R_2(t)&=\left(A-BR^{-1}B'\pi(t)\right)R_2(t)\\
&+BR^{-1}B' R_1 (T,t)'M,
% R_2(t)&=\int_0^t R_1 (t,\tau)BR^{-1}B' R_1 (T,\tau)'\mathrm{d} \tau \, M,
\end{split}
\end{equation}}
with $R_1(s,s)=I_{nk}$ \rsa{and 
$R_2(0)=0$},
where $\pi$ is the unique solution of \eqref{aux_riccati}.
We denote by $S$ the set of $k \times l$ row stochastic matrices.
% \jlnComment{Check (22). Do the 
% dimensions really work?}
For $\Lambda \in S$, define the function 
$\bar x^\Lambda: [0,T] \to \mathbb R^n$ by 
%with $\bar x^\Lambda(t) \in \mathbb R^n$ by
\begin{align} 	\label{eq: x-lambda def}
\bar x^\Lambda(t) := P_1 (R_1(t,0)\bar{X}(0)+R_2(t)\Lambda \otimes I_n \, p).
\end{align}
%and 
Next, define the finite dimensional map $F$ from $S$ into itself 
such that for all $\Lambda \in S$, 
\begin{equation} \label{finit_ope}
F(\Lambda)_{sj}=
\mathbb{P}(x_*^{s,\Lambda}(T) \in \mathcal{W}_j),
\end{equation}
where
$X^{\Lambda}_*=(x_*^{1,\Lambda},\dots, x_*^{k,\Lambda})$
%\jlnComment{changed $x_*^{s,\Lambda}$ to $x_*^{k,\Lambda}$} 
is the unique strong solution of the following SDE parameterized by $\Lambda$ 
%\jlnComment{I suspect many people will have trouble parsing the term
%$P'_\theta (R_1(.,0)\bar{X}(0)+R_2(.)\Lambda) p$}
% \iftoggle{jou}{
% \begin{align} \label{finite-operator}
% dX^\Lambda(t)&= \bigg (A X^\Lambda(t) + BU \Big(t,X^\Lambda(t),\\
% &P'_\theta \left(R_1(.,0)\bar{X}(0)+R_2(.)
% \Lambda p\right) \Big)\bigg)dt+\sigma dW(t),
% \nonumber
% \end{align}}{
% \begin{align} \label{finite-operator}
% dX^\Lambda(t)&= \left (A X^\Lambda(t) + BU \left(t,X^\Lambda(t),P'_\theta \left(R_1(.,0)\bar{X}(0)+R_2(.)
% \Lambda p\right) \right)\right)dt+\sigma dW(t),  
% \end{align}
% }
\iftoggle{jou}{
\begin{align} 
&dX^\Lambda_*(t) = \bigg( A X_*^\Lambda(t) + BU_*(t,X_*^\Lambda(t),\bar x^\Lambda) \bigg)dt
+\sigma dW(t), \nonumber \\
&\text{with } X_*^\Lambda(0) = X_*(0). 
\label{finite-operator}
\end{align}
}{
\begin{align} 
&dX_*^\Lambda(t)= \left(A X_*^\Lambda(t) + BU_* \left(t,X_*^\Lambda(t), \bar x^\Lambda \right) \right)dt+\sigma dW(t), 
&& \text{with } X_*^\Lambda(0) = X_*(0). \label{finite-operator}
\end{align}
}
The map $F$ involves the 
probability distribution of the process $X_*^{\Lambda}$. Hence, to find the value of $F(\Lambda)$, one needs to solve the Fokker-Planck equation
associated with \eqref{finite-operator}.
% \rmaComment{ Perhaps a remark to the effect that the map involves solving the Fokker-Planck equation associated with (24)}
\begin{thm} \label{lemma:existance11}
Under Assumption \ref{assum_riccati}, the following statements hold: 
\begin{enumerate}[(i)]
\item $\bar{x}$ satisfies (\ref{gen_SDE}) if and only if 
\begin{equation} \label{form_fixed}
%\bar{x}(t)=P_\theta'\left(R_1(t,0)\bar{X}(0)+R_2(t)\Lambda p\right),
\bar x = \bar x^\Lambda
\end{equation}
where $\bar x^\Lambda$ is defined in \eqref{eq: x-lambda def} and
$\Lambda$ is a fixed point of $F$.
\item %\jlnComment{Should make statement $F$ is continuous part of this (ii).}
$F$ is continuous and has at least one fixed point. 
Equivalently, (\ref{gen_SDE}) has at least one solution $\bar x$.
 \item For a uniform population, i.e, k=1, the \rsa{admissible} paths $\bar{x}$
 %\jlnComment{fixed point paths?}
 are uniformly 
 bounded with respect to $M$ and $\Lambda \in S$ for the standard $L_2$ norm
 $\Big(\int_0^T \|.\|^2\mathrm{dt}\Big)^{\frac{1}{2}}$.
\end{enumerate}
\end{thm}
%\begin{itemize}
%\item 
% Recall that point (iii) is used in Lemma \ref{contr} 
% to show that the \rma{structure of the terminal} cost \rma{is such that it incentivizes} the agents to be close to one 
% of the destination points at time $T$.
%\item 
\begin{rem}
In \cite{huang2006large, carmona2013probabilistic}, which consider
the general MFG theory, the authors show the existence and uniqueness 
of solutions for the Mckean-Vlasov equation describing the mean field 
behavior via Banach's \cite{huang2006large} %fixed point theorem 
or Schauder's fixed point theorem \cite{carmona2013probabilistic}. 
In \cite{huang2006large}, it is assumed that the optimal control law 
is regular enough (Lipschitz continuous with respect to the state 
and the distribution) in order to define a contraction, 
while in \cite{carmona2013probabilistic} the result is proved 
under the assumption of smooth and convex final cost. 
In our case, the control law \eqref{optimal-crt-exp}
%\jlnComment{Check display number.
%(5) is not the control law, but the 
%cost.} 
is not Lipschitz continuous 
with respect to $\bar{x}$. Moreover, the final cost is neither smooth 
nor convex. Hence, the Mckean-Vlasov equation \eqref{gen_SDE} might 
have multiple solutions. 
Indeed, \eqref{gen_SDE} has a number of solutions equal to the number 
of fixed points of $F$. 
%\end{itemize}
\end{rem}
Having solved the game for a continuum of players, we now return to the practical case 
of a finite population of players. Using arguments similar to those in 
\cite[Theorem 5.6]{Huang07_Large}, one can show that the MFG-based decentralized 
strategies (\ref{optimal-crt-exp}), when applied by a finite population, 
constitute an $\epsilon-$Nash equilibrium. 
\begin{thm} \label{epsilon-Nash}
The decentralized feedback strategies $u_*$ defined in (\ref{optimal-crt-exp}) 
for a fixed point path $\bar{x}$ of \eqref{gen_SDE}, when applied by $N$ players %$1,\dots,N$ of 
with dynamics (\ref{agents_dyn}), constitute an $\epsilon_N$-Nash equilibrium with respect 
to the costs (\ref{agents_cost}), where $\epsilon_N$ converges to zero as $N$ increases 
to infinity.
\end{thm}

The set of fixed points of the finite dimensional map $F$ characterizes the game 
in terms of the number of distinct $\epsilon-$Nash equilibria and the distribution 
of the choices for each of them.
In fact, Theorem \ref{lemma:existance11} establishes a one to one map between 
the solutions of the mean field equations (\ref{gen_SDE}) and the set of 
fixed point CDMs. If $\Lambda$ is a fixed point of $F$ and the players 
optimally respond to the corresponding $\bar{x}^\Lambda$ given 
by \eqref{eq: x-lambda def}, %(\ref{form_fixed}), 
then $\Lambda_{sj}$ is the fraction of agents of type $s$ that go towards $p_j$, 
and $\sum_{s=1}^k \alpha_s \Lambda_{sj}$ is the total fraction of players choosing 
this alternative.
Thus, to compute a path $\bar{x}$ satisfying (\ref{gen_SDE}), a player 
computes a fixed point $\Lambda$ of $F$ and then %implement $\Lambda$ in 
computes \eqref{eq: x-lambda def}. %(\ref{form_fixed}). 
%In the following subsection, we propose a numerical scheme to find a fixed point 
%of $F$ in the scalar binary choice case for a population with uniform dynamics 
%and cost functions, i.e., $n=m=k=1$ and $l=2$.

%\subsection{Computation of a Fixed Point of $F$: Scalar Binary Choice Case} \label{num}
%\jlnComment{You tend to have just one subsection A. in each section. This is not
%very good structure.}
%
%\jln{In this \rsa{paragraph}, 
Let us now briefly illustrate how to compute a fixed point
of $F$ in the scalar binary choice case for a population with uniform dynamics 
and cost functions, i.e., $n=m=k=1$ and $l=2$, and initial probability
density function $p_0$.
%In the binary choice case, for a population with uniform dynamics and cost functions, 
In this case, a pair $(r,1-r)$ for $r \in [0,1]$ is a fixed point of $F$ if and only 
if $r$ is a fixed point of $G$, where $G(r)=[F(r,1-r)]_1$.
Following the second point of Theorem \ref{lemma:existance11}, $F$ is continuous. As a result, $G$ is a continuous
% \jlnComment{Why (F) continuous? 
% proved in Appendix B?
% Add the Thm 5 (ii) the fact that F 
% is continuous.} 
function from $[0,1]$ into itself. 
Thus, we can apply the bisection method to $G(r)-r$
to find a fixed point of $G$, %this function, 
if we can compute the value of $G$ at any $r \in [0,1]$. 
But $G(r) = \int_{-\infty}^c p_r(T, x) dx$, where %, for $r \in [0,1]$, 
%is the value at 
$c=(p_1+p_2)/2$ %of the cumulative distribution function 
and $p_r(t,x)$ is the probability density of $X^{(r,1-r)}(t)$
%of $X^{(r,1-r)}(T)$ 
defined by (\ref{finite-operator}), which can be computed numerically  
%by propagating 
by solving the following Fokker-Planck equation associated to \eqref{finite-operator}, 
e.g., via an implicit finite difference scheme \cite{pichler2013numerical}
\begin{align} \label{FP-eq}
\frac{\partial p_r(t,x)}{\partial t} =
-\frac{\partial \left (
\mu (t,x,r)p_r(t,x) 
\right )}{\partial x} +
\frac{\sigma^2}{2} 
\frac{\partial^2 p_r(t,x)}{\partial x^2}, 
\end{align}
with $p_r(0,x)=p_0(x), \forall x \in \mathbb{R}$. 
%\jlnComment{was $p_0$ ever defined?}
Here 
%$\mu (t,x,r) = Ax+Bu_* \left(t,x,R_1(t,0)\bar{x}(0)+R_2(t) p_{(r,1-r)}\right)$.
$\mu (t,x,r) = Ax+Bu_* \left(t,x,\bar x^{(r,1-r)}\right)$ for 
$\bar x^{(r,1-r)}(t) = R_1(t,0)\bar{x}(0)+R_2(t) \left(rp_1+(1-r)p_2\right)$,
see \eqref{eq: x-lambda def}.
\section{Simulation Results} \label{sim}

\begin{figure}[htb]
    \centering
\includegraphics[width=0.49\textwidth]{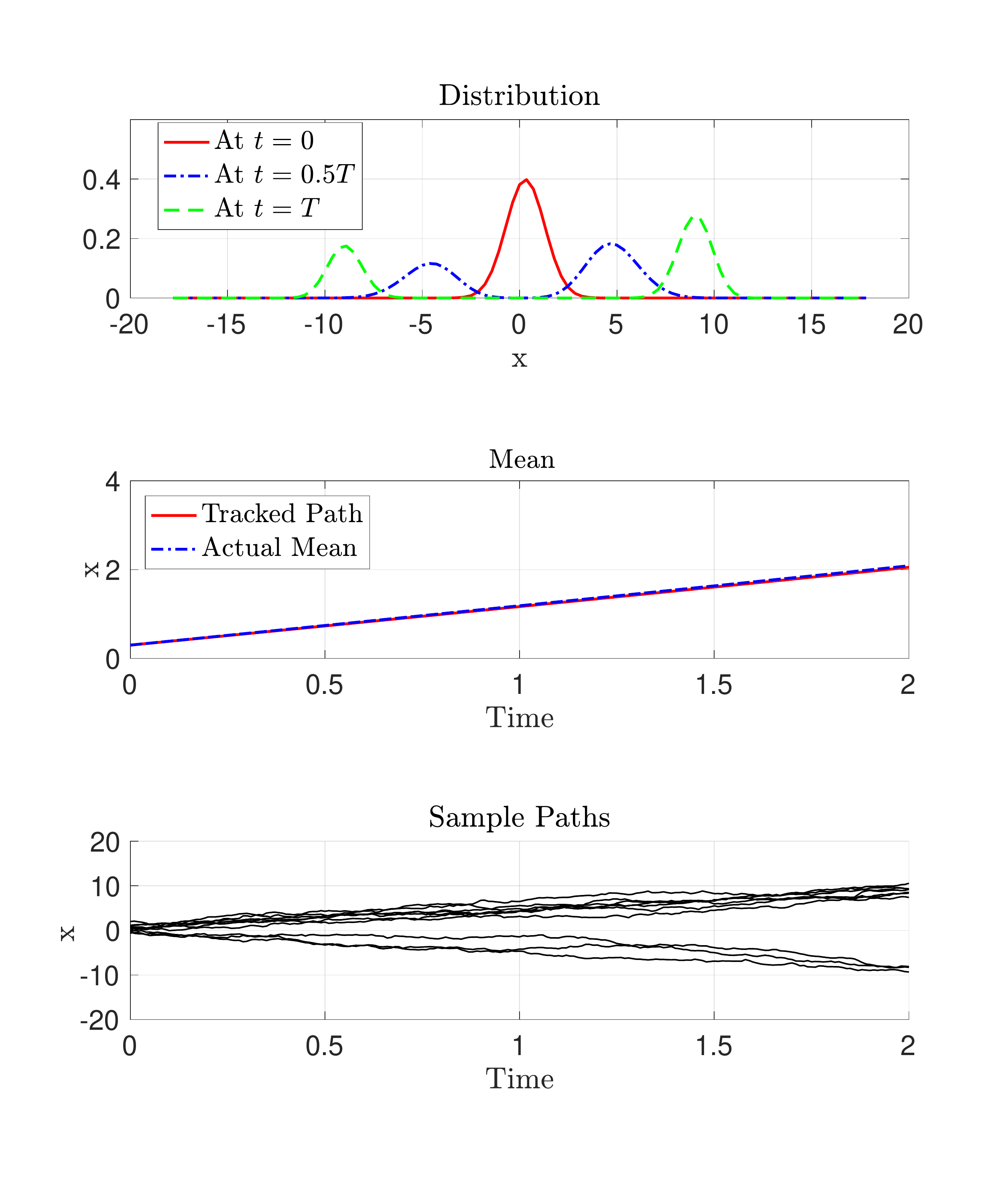}  %width=0.71\textwidth
    \caption{Distribution evolution, mean trajectory, tracked path and sample paths for a weak social effect ($Q=0.1$). Here $r=0.39$.
    }
    \label{Fig.1}
\end{figure}

To illustrate the dynamics of our collective decision mechanism, 
we consider a group of agents with parameters $A=0.1$, 
$B=0.2$, $R=5$, $M=500$, $T=2$ and $\sigma=1.5$. 
The agents, initially drawn from the normal distribution
$\mathcal{N}(0.3,1)$, choose between the alternatives
$p_1=-10$ and $p_2=10$. At first, we consider a weak social effect ($Q=0.1$). 
Following the numerical scheme at the end of Section \ref{section-MFG-equ}, we find
a fixed point $r=0.39$. Accordingly, a player applying its
decentralized MFG-based strategy is at time $T$ closer to $p_1$ with 
probability $0.39$. Equivalently, if we draw independently 
from the initial distribution a large number of players with independent 
Wiener processes, then the percentage of players that will be 
at time $T$ closer to $p_1$ converges to $39\%$ as the size of the population 
increases to infinity. Thus, the majority of the players \rma{choose} $p_2$.
Figure \ref{Fig.1} shows the distribution at time $t=0$, $t=0.5T$ and $t=T$, the
mean of a generic agent, the tracked path (admissible path (\ref{form_fixed})), and the sample paths of $10$
players choosing between $-10$ and $10$ under the weak social effect. As shown 
in this figure, the mean replicates the tracked path computed using 
the fixed point $r=0.39$ and (\ref{form_fixed}). 

%  \begin{figure}[H]
%     \centering
% \includegraphics[width=0.5\textwidth]{figures/q0_sim.pdf} %width=0.65\textwidth
%     \caption{Sample paths of the agents for a weak social effect ($Q=0.1$).}
%     \label{Fig.2}
% \end{figure}

\begin{figure}[htb]
    \centering
\includegraphics[width=0.5\textwidth]{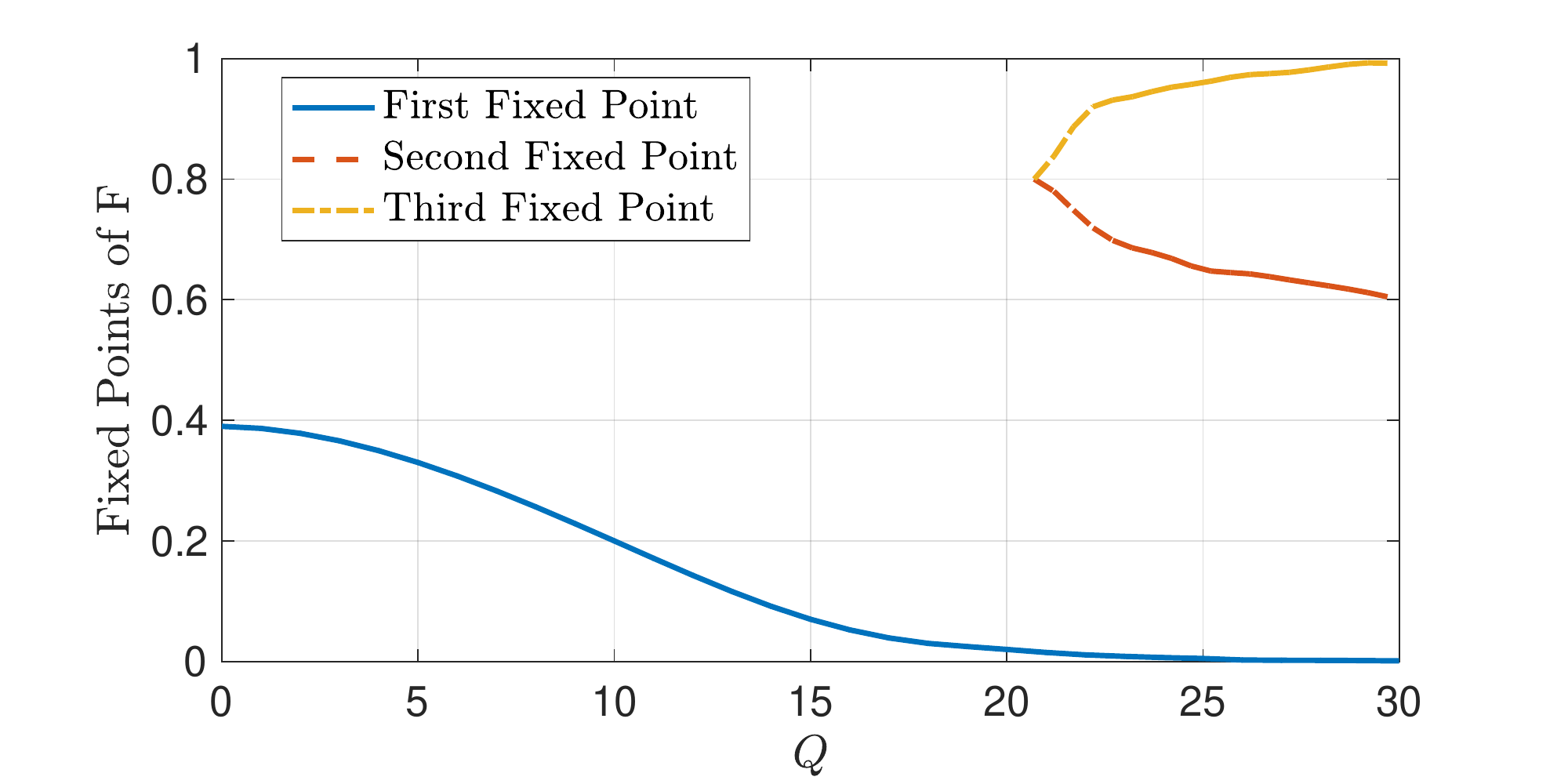}
    \caption{Influence of the coefficient $Q$ on the multiplicity of the fixed
    points}
    \label{Fig.4}
\end{figure}

\begin{figure}[htb]
    \centering
\includegraphics[width=0.5\textwidth]{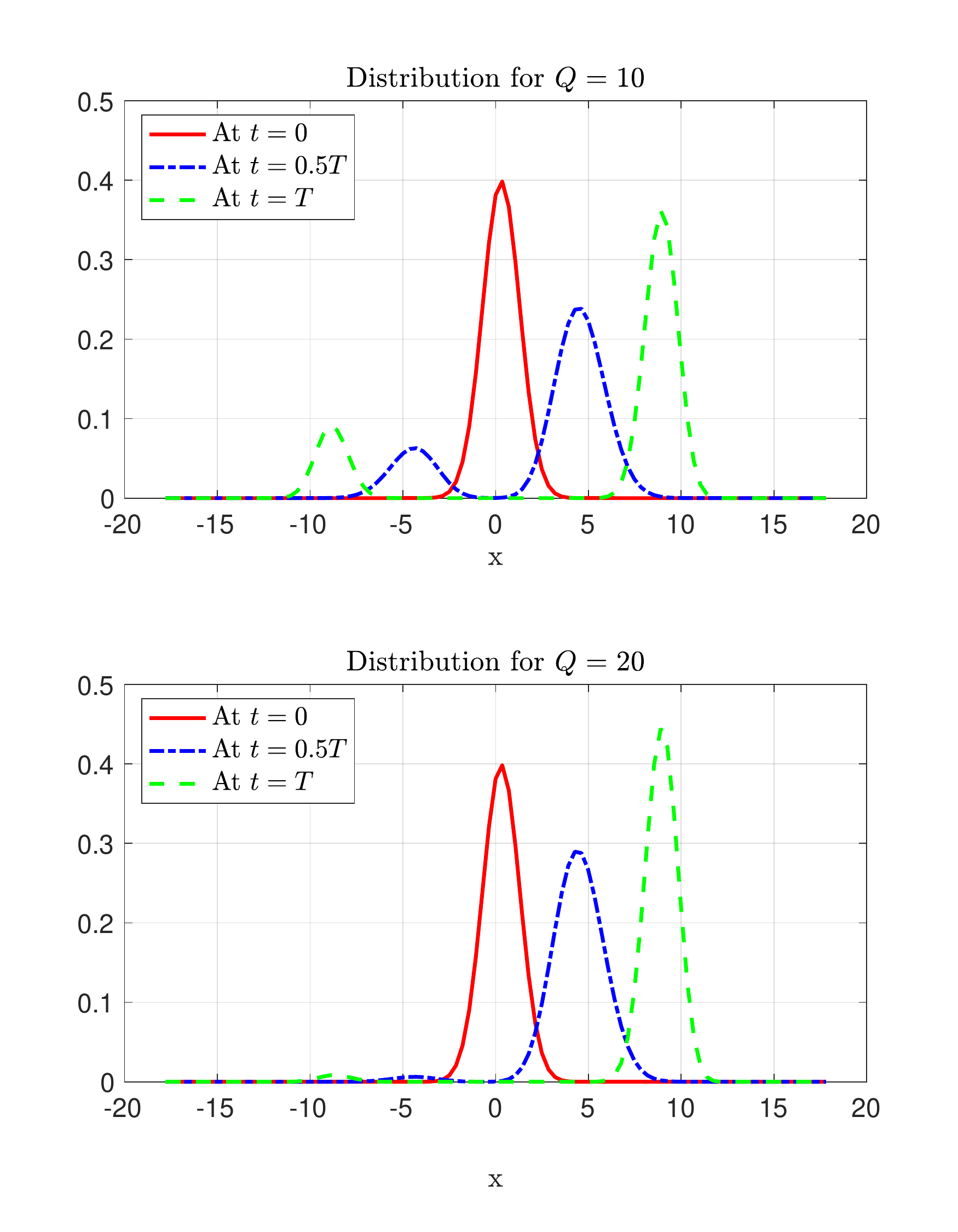}
    \caption{Influence of the social effect on the distribution of the agents. Distribution evolution for medium ($Q=10$) and strong ($Q=20$) social effects. In the first case, $r=0.2$, while in second case, $r=0.02$.}
    \label{Fig.2}
\end{figure}

Figure \ref{Fig.4} shows that for \rma{sufficiently} small values of $Q$ ($Q<21$) the fixed points 
of $F$ are unique. When the social 
effect $Q$ exceeds $21$, $F$ has 
three fixed points, where two of them correspond to consensus on one alternative. Indeed as $Q$ increases arbitrarily, agents essentially forget temporarily about the final cost, and the problem becomes a classical rendez-vous MFG where they tend to merge towards each other rapidly. If this occurs around the middle of the destinations segment, then this is clearly an unstable situation where most of the time, they end up splitting classically according to initial conditions; however, some large deviations are possible whereby a significant fraction decides to choose one destination, thus pulling everyone else towards it, which \rma{may help explain} the non uniqueness of outcomes.
Figure \ref{Fig.2} illustrates the evolution
of the distributions that correspond to the first fixed point with $Q=10$ and $Q=20$.
% \jlnComment{Note also that as $Q$ becomes large, two of the fixed points
% correspond to consensus on one alternative.}

%  \begin{figure}[H]
%     \centering
% \includegraphics[width=0.5\textwidth]{figures/q10_sim.pdf}
%     \caption{Sample Paths for a medium social effect ($Q=10$).}
%     \label{Fig.4}
% \end{figure}
% \begin{figure}[H]
%     \centering
% \includegraphics[width=0.5\textwidth]{figures/Figure3.pdf}
%     \caption{Distribution evolution for a strong 
%     social effect ($Q=20$). Here $r=0.02$.}
%     \label{Fig.5}
% \end{figure}

%  \begin{figure}[H]
%     \centering
% \includegraphics[width=0.5\textwidth]{figures/q20_sim.pdf}
%     \caption{Sample paths for a strong social effect ($Q=20$).}
%     \label{Fig.6}
% \end{figure}

\begin{figure}[htb]
    \centering
\includegraphics[width=0.5\textwidth]{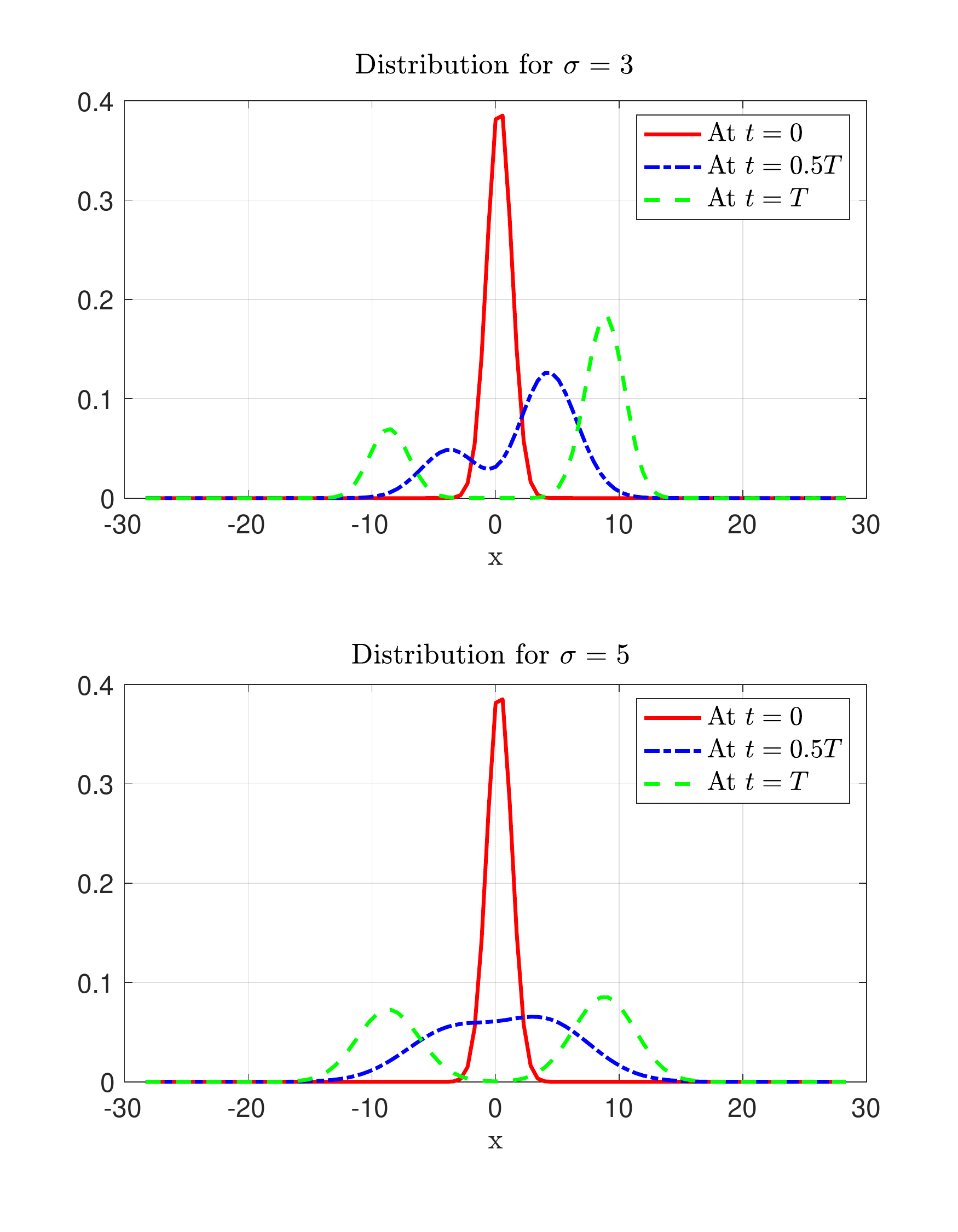}
    \caption{Influence of the noise on the distribution of the agents. Distribution evolution for $\sigma=3$ and $\sigma=5$. In the first case, $r=0.28$, while in the second case, $r=0.46$.}
    \label{Fig.3}
\end{figure}

To illustrate the effect of the noise intensity on the behavior of the group, 
we fix $Q = 20$ and we increase $\sigma$ from $1.5$ to $5$. For 
$\sigma=1.5$, $r=0.02$ (Figure \ref{Fig.2}), 
but for $\sigma=3$, $r$ increases to $0.28$ (Figure \ref{Fig.3}) and for $\sigma=5$, $r=0.46$ (Figure \ref{Fig.3}). Thus, the higher the noise
the more evenly distributed the players are  between the alternatives.

% \begin{figure}[H]
%     \centering
% \includegraphics[width=0.5\textwidth]{figures/sigma3_sim.pdf}
%     \caption{Noise Intensity Effect - $\sigma=3$ - Sample Paths}
%     \label{Fig.8}
% \end{figure}
% \begin{figure}[H]
%     \centering
% \includegraphics[width=0.6\textwidth]{figures/sigma5_dist.pdf}
%     \caption{Noise Intensity Effect - $\sigma=5$ - Distribution Evolution,
%     Mean Trajectory and Tracked path}
%     \label{Fig.9}
% \end{figure}
% \begin{figure}[H]
%     \centering
% \includegraphics[width=0.5\textwidth]{figures/sigma5_sim.pdf}
%     \caption{Noise Intensity Effect - $\sigma=5$ - Sample Paths}
%     \label{Fig.10}
% \end{figure}
\section{Conclusion}\label{conclusion} 

We study within the framework of mean field game theory a dynamic 
collective choice model with social interactions. 
We introduce the Min-LQG optimal control problem and give an explicit 
form of a generic agent's best response (Min-LQG optimal control
law). The Min-LQG problem can be interpreted at each time step 
as a static discrete choice model where the cost of choosing 
one of the alternatives has an additional term that increases with 
the risk of being driven by the process noise to the other  %Brownian motion
alternatives. We show the existence of closed loop decentralized 
$\epsilon-$Nash strategies. Moreover, we characterize these strategies 
by a CDM describing the way the 
population splits between the alternatives, which is a 
fixed point of a well defined finite dimensional map.
\appendices  

\section{}  \label{ProofA}
In this appendix, we provide the proofs of lemmas and theorems 
related to a generic agent's best response.
% \subsection*{Proof of Lemma \ref{spe-case}} \label{Proof1}
% The unique solution of (\ref{Heat_equation}) is 
% \cite[Section 2.3, Theorem 1]{evans1998partial}:  %\jlnComment{Give ref. to specific Theorem in that book}
% \begin{equation} \label{proof-psi}
% \begin{split}
% \psi (t,x) =\frac{1}{\sigma \sqrt{2\pi(T-t)}} 
% \int_{-\infty}^{\infty}&  \exp \left (
% -\frac{(x-y)^2}{2\sigma^2 (T-t)} \right)
% \psi(T,y)dy\\ 
% =\frac{1}{\sigma \sqrt{2\pi(T-t)}} 
% \Bigg \{
% &\int_{-\infty}^c  \exp \left (
% -\frac{(x-y)^2}{2\sigma^2 (T-t)} -
% \frac{MB^2}{2\sigma^2R} 
% (y-p_1)^2 \right)dy \\
% &+\int^{\infty}_c  \exp \left (
% -\frac{(x-y)^2}{2\sigma^2 (T-t)} -
% \frac{MB^2}{2\sigma^2R} 
% (y-p_2)^2 \right)dy \Bigg\}.
% \end{split}
% \end{equation}
% By expanding the integrands in (\ref{proof-psi}) and completing the squares, we get for $j=1,2$,
% \begin{align*}
% \frac{(x-y)^2}{2\sigma^2 (T-t)} +
% \frac{MB^2}{2\sigma^2R} 
% (y-p_j)^2&= 
% \frac{B^2}{2\sigma^2R}\Pi(t)x^2
% +\frac{B^2}{\sigma^2R}\beta_j(t)x
% +\frac{MB^2p_j^2}{2\sigma^2(R+MB^2(T-t))}
% \\
% &+\frac{R+MB^2(T-t)}{2\sigma^2R(T-t)}
% \left( y- \frac{xR+MB^2p_j(T-t)}{R+MB^2(T-t)} \right)^2,
% \end{align*}
% \rsa{where $\Pi$ and $\beta_j$ are defined in 
% \eqref{pi-beta-delta}.}
% \jln{Using these expressions in the integrands of \eqref{proof-psi} and  
% making a change of variable}  
% \[z_j= \sqrt{\frac{R+MB^2(T-t)}{R}} 
% \left (y- \frac{xR+MB^2p_j(T-t)}{R+MB^2(T-t)} \right),\] 
% for $j=1,2$, we get 
% (\ref{sol_heat}).
\subsection*{Proof of Theorem \ref{thm-hjb}} \label{Proof4}

%Before proving Theorem \ref{thm-hjb}, we need \jln{some preliminary results}. %the following results. 
%\jln{First}, Lemma \ref{lemma_g} asserts that if the mean of an indexed normal 
%distribution converges to some value, and the variance shrinks to zero, then the probability of a 
%closed subset that does not include the limit of the mean converges to zero.
%\jlnComment{Can't this paragraph and lemma be restated in terms of cv in MS implies cv in proba?}
We start with a technical result on the mean-square convergence of random variables.
\begin{lem} \label{lemma_g}
%\jlnComment{I think it would be clearer to state such a result (and others) 
%in terms of probability and random variables rather than integrals and calculus.
%This is essentially a result related to convergence in mean square 
%implying convergence in probability...}
Let $I$ be a closed subset of $\mathbb{R}^n$. 
Let $X_k \in \mathbb{R}^n$ be a sequence of random variables with
finite first and second moments. If $\mathbb E[X_k] =: \mu_k \to \mu$ for some
vector $\mu$ not in $I$, and $\mathbb E[\|X_k - \mu_k\|^2 ] \to 0$, then
%distributions $\mathcal{N}(\mu_k,\Sigma_k)$. \jln{If $\mu_k$} % on some probability space
%converges to $\mu \not\in I$ and $\Sigma_k$ converges to $0$, then
$
\lim\limits_{k \rightarrow \infty} \mathbb{P}\left(X_k \in I \right)=0.
$
\end{lem}

\begin{IEEEproof}
$I \subset \mathbb{R}^n$ is a closed set and $\mu \notin I$, so the
distance $d$ between $\mu$ and $I$ is strictly positive. Since, $\mu_k$ 
converges to $\mu$, there exists $k_0>0$ such that for all $k \geq 0$ and for all
$x$ in $I$ we have $\|x-\mu_k\| \geq d/2$. 
Hence, using Chevyshev's inequality \cite[Theorem 1.6.4]{durrett2010probability},
\[
\mathbb{P} \left( X_k  \in I\right) \leq \mathbb{P} \left( \|X_k-\mu_k\| \geq d/2 \right)
\leq \frac{4}{d^2} \mathbb{E} [\|X_k-\mu_k\|^2],
\]
for all $k \geq k_0$. The result follows since the right-hand side of the inequality
is assumed to converge to $0$.
%
% and $\epsilon>0$ such that $\|X_k-\mu_k\|>\epsilon$, 
% for all $k>k_0$, for all $X_k \in I$. Hence, for all $k>k_0$,
% % \begin{align*}
% % &\frac{1}{\sqrt{|2\pi\Sigma_{t_k}|}}
% % \int_I \exp \left ( 
% % -\left\|y-f(t_k,x_k) 
% % \right\|^2_{\Sigma_{t_k}^{-1}} \right )dy \\
% % &\leq \frac{1}{\sqrt{|2\pi\Sigma_{t_k}|}}
% % \int_{\|y-f(t_k,x_k)\|>\epsilon} \exp \left ( 
% % -\left\|y-f(t_k,x_k) 
% % \right\|^2_{\Sigma_{t_k}^{-1}} \right )dy 
% % \leq \frac{\text{Tr} (\Sigma_{t_k})}{\epsilon^2},
% % \end{align*}
% \[\mathbb{P}\left( X_k  \in I\right) \leq \mathbb{P}\left( \|X_k-\mu_k\|>\epsilon \right)
% \leq \frac{1}{\epsilon^2} \mathbb{E}\|X_k-\mu_k\|^2 = 
% \frac{1}{\epsilon^2} \text{Tr} (\Sigma_k),\]
% \jln{where the second inequality is Chebyshev's inequality \cite[Theorem 1.6.4]{durrett2010probability}}.
% %The last inequality is \jln{\st{a result of}} Chebyshev inequality. 
% %But, $\text{Tr} (\Sigma_{t_k})$ converges to zero as $k$ goes to infinity. \jln{\st{Hence the result.}}
% \jln{The result follows from the fact that the variance 
% $\Sigma_k$ of $X_k$ goes to zero as $k$ goes to infinity.}
\end{IEEEproof}

%\subsection*{Proof of Lemma \ref{lem-cont}} \label{Proof3}  % JLN: currently here

The following lemma concerns the regularity of the solution provided in Theorem \ref{thm-hjb}. 
\begin{lem} \label{lem-cont} 
$V$ defined in \eqref{sol_hjb} is in $C^{1,2}([0,T)\times \mathbb{R}^n) \cap C([0,T] \times \mathbb{R}^n)$.
\end{lem}

\begin{IEEEproof}
Note that for $g_j$ defined in \eqref{g1_g2}, we can write
\iftoggle{jou}{
\begin{equation} \label{g1_g2 bis}
\begin{split}
g_j(t,x)&=\mathbb{P}\left (x^{(j)}(T) \in \mathcal{W}_j \Big| x^{(j)}(t)=x\right)\\
&=\frac{1}{\sqrt{|2\pi\Sigma_t|}} \int_{\mathcal{W}_j} \exp 
\bigg( -\Big\|y-\alpha(T,t)x
\\
&+\int_t^T \alpha(T,\tau)BR^{-1}B'\beta_j(\tau)d\tau \Big\|^2_{\Sigma^{-1}_t} \bigg) dy\\
\text{for } \Sigma_t&= \int_t^T \alpha(T,\tau)\sigma \sigma' \alpha(T,\tau)' d\tau,
\end{split}
\end{equation}}{
\begin{equation} \label{g1_g2 bis}
\begin{split}
g_j(t,x)&=\mathbb{P}\left (x^{(j)}(T) \in \mathcal{W}_j \Big| x^{(j)}(t)=x\right)\\
&=\frac{1}{\sqrt{|2\pi\Sigma_t|}} \int_{\mathcal{W}_j} \exp 
\left( -\left\|y-\alpha(T,t)x
+\int_t^T \alpha(T,\tau)BR^{-1}B'\beta_j(\tau)d\tau \right\|^2_{\Sigma^{-1}_t} \right) dy\\
\text{for } \Sigma_t&= \int_t^T \alpha(T,\tau)\sigma \sigma' \alpha(T,\tau)' d\tau,
\end{split}
\end{equation}}
where $x^{(j)}$, $\Pi$ and $\beta_j$
are defined in \eqref{aux_st} and \eqref{pi-beta-delta}, 
and the matrix-valued function $\alpha(t,s)$ is the unique solution of 
\begin{equation}	\label{eq: definition of alpha}
\frac{d}{dt}\alpha(t,s)=\left(A-BR^{-1}B'\Pi(t)\right)\alpha(t,s),
\end{equation}
with $\alpha(s,s)=I_n$. The expression \eqref{g1_g2 bis} follows from the fact that the solution of a linear SDE
with deterministic initial condition has a normal distribution \cite[Section 2.5]{karatzas2012brownian}.
In view of (\ref{sol_hjb}), (\ref{g1_g2 bis}), and $\Sigma_t \succ 0$ for all $t \in [0,T)$, 
$V$ is in $C^{1,2}([0,T)\times \mathbb{R}^n)$. It remains to show the continuity on $\{T\}\times \mathbb{R}^n$. We start by considering 
$x \in \mathbb{R}^n \setminus \cup_{j=1}^l \partial \mathcal{W}_{j}$, and $(t_k,x_k) \in [0,T) \times \mathbb{R}^n$ converging to $(T,x)$. We have $x \in \text{Int}(\mathcal{W}_{j_0})$ for some $j_0 \in \{1,\dots,l\}$, and $x \not\in \mathcal{W}_{j}$ for $j \neq j_0$. In view of \eqref{g1_g2 bis}, $g_j(t_k,x_k)$ is the probability that a Gaussian vector of mean $\alpha(T,t_k)x_k
-\int_{t_k}^T \alpha(T,\tau)BR^{-1}B'\beta_j(\tau)d\tau$ (which converges to $x$ with $k$) and covariance $\Sigma_{t_k}$ (which converges to $0$ with $k$) is in  the closed set $\mathcal{W}_j$.  In this way, each $j$  defines a distinct sequence of random variables associated with the $(t_k,x_k)$'s. Now if one considers the closed set $I$ of  Lemma \ref{lemma_g} to be any of the closed sets  $\mathcal{W}_j$'s for  $j \neq j_0$   , one can conclude from this lemma that $g_j(t_k,x_k)$ must converge  to $0$ for $j \neq j_0$ and, as a consequence, to $1$ for $j=j_0$ since the  $\mathcal{W}_j$'s form a partition of the state space. Therefore, $V(t_k,x_k)$ converges to $V(T,x)$.
Thus, $V$ is continuous on $[0,T] \times \left(\mathbb{R}^n \setminus \cup_{j=1}^l \partial \mathcal{W}_{j}\right).$ 
Finally, consider a sequence $(t_k,x_k) \in [0,T) \times \mathbb{R}^n$ converging to
$(T,c)$, with $c \in \cup_{j=1}^l \partial \mathcal{W}_{j}$. We show that $V(t_k,x_k)$
converges to $V(T,c)$. 
Up to renumbering the Voronoi cells, we can assume without loss of generality 
that $c \in \partial \mathcal{W}_{j}$ for all $j$ in $\{1,\ldots,z\}$ and 
%$c \in \cap_{j=1}^z \partial VC_{j}$, and
$c \notin \cup_{j=z+1}^l \mathcal{W}_{j}$, for some $1\leq z\leq l$. We have,
\iftoggle{jou}{
\begin{align*}
I_0&=\sum_{j=1}^l\exp \left( -\eta V_j(t_k,x_k) \right)
g_j(t_k,x_k)\\
&=\sum_{j=1}^z\exp \left( -\eta V_j(t_k,x_k) \right)
g_j(t_k,x_k)\\
&+\sum_{j=z+1}^l\exp \left( -\eta V_j(t_k,x_k) \right)
g_j(t_k,x_k).
%\\
%&\exp \left( -
%\frac{b^2}{\sigma^2r} V_2(t_k,x_k) \right)
%+I_1(t_k,x_k,\delta)+I_2(t_k,x_k,\delta),
\end{align*}}{
\begin{align*}
I_0&=\sum_{j=1}^l\exp \left( -\eta V_j(t_k,x_k) \right)
g_j(t_k,x_k)\\
&=\sum_{j=1}^z\exp \left( -\eta V_j(t_k,x_k) \right)
g_j(t_k,x_k)+\sum_{j=z+1}^l\exp \left( -\eta V_j(t_k,x_k) \right)
g_j(t_k,x_k).
%\\
%&\exp \left( -
%\frac{b^2}{\sigma^2r} V_2(t_k,x_k) \right)
%+I_1(t_k,x_k,\delta)+I_2(t_k,x_k,\delta),
\end{align*}
}
Since $c \notin \cup_{j=z+1}^l \mathcal{W}_{j}$, one can use an argument similar to that above to show that 
the second term of the right-hand side of the second equality converges to $0$. 

Next, let $\epsilon>0$ and fix $r >0$ small enough so that $\bar{B}(c,r) \subset \left(\cap_{j=z+1}^l \mathcal{W}_j\right)^C$.
%\jlnComment{$k$ should be $z$?} 
The value of $r$ will be determined later. 
The first term can be written
\begin{align*}
&\sum_{j=1}^z\exp \left( -\eta V_j(t_k,x_k) \right)
g_j(t_k,x_k)=I_1+I_2,
\end{align*}
where
\iftoggle{jou}{
\begin{align*}
I_1=&\sum_{j=1}^z 
\exp \left( -\eta V_j(t_k,x_k) \right)\\
&
\times \mathbb{P}\left ( 
x^{(j)}(T) \in \mathcal{W}_j \cap \bar{B}(c,r) \Big| x^{(j)}(t_k)=x_k\right)\\
I_2=&\sum_{j=1}^z 
\exp \left( -\eta V_j(t_k,x_k) \right)\\
&
\times \mathbb{P}\left ( 
x^{(j)}(T) \in \mathcal{W}_j \setminus \bar{B}(c,r) \Big| x^{(j)}(t_k)=x_k\right).
\end{align*}}{
\begin{align*}
I_1=&\sum_{j=1}^z 
\exp \left( -\eta V_j(t_k,x_k) \right)
\mathbb{P}\left ( 
x^{(j)}(T) \in \mathcal{W}_j \cap \bar{B}(c,r) \Big| x^{(j)}(t_k)=x_k\right)\\
I_2=&\sum_{j=1}^z 
\exp \left( -\eta V_j(t_k,x_k) \right)
\mathbb{P}\left ( 
x^{(j)}(T) \in \mathcal{W}_j \setminus \bar{B}(c,r) \Big| x^{(j)}(t_k)=x_k\right).
\end{align*}
}
% In view of $\lim\limits_{k \rightarrow \infty}\Sigma_{t_k}=0$, 
% the Gaussian distribution defined by $\mathbb{P}\left ( 
% x^{(j)}(T) \in \mathcal{A} \Big| x^{(j)}(t_k)=x_k\right)$, 
% for all Borel sets $\mathcal{A}$ converges in distribution\jlnComment{rewrite sentence} 
% to the Dirac distribution $\delta_c$. But 
% $\{VC_j \cap \bar{B}(c,r)\}_{j=1}^z$
% partition $\bar{B}(c,r)$ (up to 
% a \jln{\st{Lebesgue zero measure subset}} \jln{subset of Lebesgue measure zero}). Hence,
% \[
% \lim\limits_{k \rightarrow \infty}
% \sum_{j=1}^z 
% \mathbb{P}\left ( 
% x^{(j)}(T) \in VC_j \cap \bar{B}(c,r) \Big| x^{(j)}(t_k)=x_k\right)=1.
% \]
% Moreover, $\lim\limits_{k \rightarrow \infty} V_j(t_k,x_k)=V(T,c)$, for all $j=1,\dots,z$. 
% Therefore, we obtain that $I_1$ converges to $\exp(-\eta V(T,c))$, and the result follows.
By Lemma \ref{lemma_g},
$I_2$
converges to zero. Next, by solving the linear differential equations in \eqref{pi-beta-delta} and replacing the expressions of $\beta_j$ and $\delta_j$ in the expression \eqref{g1_g2 bis} of $g_j$, one can show
that under Assumption \ref{assumption_linear}
\iftoggle{jou}{
\begin{align*}
I_1&=\exp \left( -\eta V_0(t_k,x_k) \right)\\
&\times \sum_{j=1}^z 
\int_{\mathcal{W}_j \cap \bar{B}(c,r)} f_k(y) \exp\left(\eta(\|y\|_M^2-\|y-p_j\|_M^2)\right) dy,
\end{align*}}{
\begin{align*}
I_1&=\exp \left( -\eta V_0(t_k,x_k) \right)\sum_{j=1}^z 
\int_{\mathcal{W}_j \cap \bar{B}(c,r)} f_k(y) \exp\left(\eta(\|y\|_M^2-\|y-p_j\|_M^2)\right) dy,
\end{align*}
}
where $f_k(y)$ is the probability density function 
%\jlnComment{cumulative distribution 
%function? probability density 
%function?}
of the Gaussian distribution of mean $\alpha(T,t_k)x_k
-\int_{t_k}^T \alpha(T,\tau)BR^{-1}B'\beta_0(\tau)d\tau$ and variance $\Sigma_{t_k}$, and $V_0$ and $\beta_0$ are equal to 
$V_j$ and $\beta_j$ defined in \eqref{aux_cost}-\eqref{pi-beta-delta} but for $p_j=0$. By the definition of $c$, $\|c-p_1\|_M^2=\dots=\|c-p_z\|_M^2$. Hence,
\iftoggle{jou}{
\begin{align*}
&I_1=\exp \left( -\eta (V_0(t_k,x_k)-\|c\|_M^2+\|c-p_j\|_M^2) \right)\\
&\times \sum_{j=1}^z 
 \int_{\mathcal{W}_j \cap \bar{B}(c,r)} f_k(y)dy\\
&+\exp \left( -\eta V_0(t_k,x_k\right))\sum_{j=1}^z 
\int_{\mathcal{W}_j \cap \bar{B}(c,r)} f_k(y)f(y)dy \triangleq I_3+I_4
\end{align*}}{
\begin{align*}
I_1&=\exp \left( -\eta (V_0(t_k,x_k)-\|c\|_M^2+\|c-p_j\|_M^2) \right)\sum_{j=1}^z 
 \int_{\mathcal{W}_j \cap \bar{B}(c,r)} f_k(y)dy\\
&+\exp \left( -\eta V_0(t_k,x_k\right))\sum_{j=1}^z 
\int_{\mathcal{W}_j \cap \bar{B}(c,r)} f_k(y)f(y)dy\\
& \triangleq I_3+I_4,
\end{align*}
}
where $f(y)=\exp \left( \eta (\|y\|_M^2-\|y-p_j\|_M^2) \right)-\exp \left( \eta (\|c\|_M^2-\|c-p_j\|_M^2) \right)$. $V_0(t_k,x_k)$ converges to $V_0(T,c)=\|c\|_M^2$ 
%\jlnComment{sign error?},
$f_k$ converges in distribution to a point mass at $c$, and $\mathcal{W}_j \cap \bar{B}(0,c)$, $j=1,\dots,z$, is a partition of $\bar{B}(0,c)$. Therefore, $I_3$
converges to $\exp(-\eta\|c-p_j\|_M^2)=\exp(-\eta V(T,c))$. $f$ is continuous, and $f(c)=0$. Hence, one can choose $r$
small enough so that $|f(y)|<\epsilon$
for all $y \in \bar{B}(c,r)$. Thus,
$|I_4| \leq \epsilon$, and 
$\limsup\limits_k |I_0-\exp(-\eta V(T,c))| \leq \epsilon$.
%\jlnComment{I_1?}
Since $\epsilon$ is arbitrary, $I_0$ converges to $\exp(-\eta V(T,c))$. 
This proves the result.
\end{IEEEproof}

%\subsection*{Proof of Theorem \ref{thm-hjb}} \label{Proof4}
To finish the proof of Theorem \ref{thm-hjb},
%In view of Lemma \ref{lem-cont}, 
it remains to show that $V$ satisfies the HJB equation (\ref{general_HJB}).  
We define the transformations by a generalized Hopf-Cole transformation \cite[Chapter 4-Section 4.4]{evans1998partial} of 
$V_j(t,x)$, $\psi_j (t,x) = \exp \left ( -\eta
V_j(t,x)
\right )$, for $j=1,\dots,l$. Recall \cite[Chapter 6]{yong1999stochastic} that the optimal cost-to-go
$V_j$ satisfies the HJB equation \eqref{general_HJB}, but with the boundary condition equal to $V_j(T,x)=\|x-p_j\|_M^2$. By multiplying the right-hand and left-hand sides of \eqref{general_HJB} by 
$-\eta\exp \left ( -\eta
V_j (t,x)
\right )$, one obtain that
\begin{align*}
&-\frac{\partial \psi_j }{\partial t} 
= 
x'A'\frac{\partial \psi_j}{\partial x}+\frac{1}{2} \text{Tr} \left( \sigma' 
\frac{\partial^2 \psi_j}{\partial x^2} \sigma \right)
-\eta\|x-\bar{x}\|^2_Q
\psi_j \\
&+\eta\exp \left ( -\eta
V_j (t,x)
\right )\frac{1}{2}\left( 
\frac{\partial V_j}{\partial x} \right)'\left( BR^{-1}B'-\eta\sigma \sigma'\right) \frac{\partial V_j}{\partial x}.
\end{align*}
Thus, under Assumption \ref{assumption_linear}, we get
\begin{align} \label{parab_equation}
-\frac{\partial \psi_j }{\partial t} 
&= 
x'A'\frac{\partial \psi_j}{\partial x}+\frac{1}{2} \text{Tr} \left( \sigma' 
\frac{\partial^2 \psi_j}{\partial x^2} \sigma \right)
-\eta\|x-\bar{x}\|^2_Q
\psi_j\nonumber\\
\psi_j (T,x) &= \exp \left ( - \eta \|x-p_j\|^2_M \right), 
\;\; \forall x \in \mathbb R^n.
\end{align}
Define $\psi (t,x) = \exp \left ( -\eta V (t,x)
\right )$ the transformation of $V(t,x)$ defined in (\ref{sol_hjb}). Hence, we have
$
\psi(t,x)=\sum_{j=1}^l\psi_j(t,x)g_j(t,x).
$
Equation (\ref{parab_equation}), Assumption \ref{assumption_linear} and the
identity 
$\frac{\partial \psi_j}{\partial x}=
-\eta\left(\Pi x +\beta_j \right)\psi_j$, where $\Pi$ and 
$\beta_j$ are defined in \eqref{pi-beta-delta}, imply 
\iftoggle{jou}{
\begin{align*}
&\frac{\partial \psi }{\partial t} 
+ 
x'A'\frac{\partial \psi}{\partial x}+\frac{1}{2} \text{Tr} \left( \sigma' 
\frac{\partial^2 \psi}{\partial x^2} \sigma \right)
-\eta\|x-\bar{x}\|^2_Q
\psi \\ 
&=  
\sum_{j=1}^l \Bigg ( \frac{\partial g_j}{\partial t} 
+ \left(Ax-BR^{-1}B'\Pi x - BR^{-1}B'
\beta_j\right)'\frac{\partial g_j}{\partial x}
\\
&+\frac{1}{2}\text{Tr} \left( \sigma' 
\frac{\partial^2 g_j}{\partial x^2} \sigma \right)\Bigg) \psi_j.
\end{align*}}{
\begin{align*}
&\frac{\partial \psi }{\partial t} 
+ 
x'A'\frac{\partial \psi}{\partial x}+\frac{1}{2} \text{Tr} \left( \sigma' 
\frac{\partial^2 \psi}{\partial x^2} \sigma \right)
-\eta\|x-\bar{x}\|^2_Q
\psi \\ 
&=  
\sum_{j=1}^l \left ( \frac{\partial g_j}{\partial t} 
+ \left(Ax-BR^{-1}B'\Pi x - BR^{-1}B'
\beta_j\right)'\frac{\partial g_j}{\partial x}
+\frac{1}{2}\text{Tr} \left( \sigma' 
\frac{\partial^2 g_j}{\partial x^2} \sigma \right)\right) \psi_j.
\end{align*}
}
The process $x^{(j)}$ satisfies the SDE (\ref{aux_st}). 
Therefore, by Kolmogorov's backward equation \cite[Section 5.B]{karatzas2012brownian}, 
\iftoggle{jou}{
\begin{align*}
&\frac{\partial g_j}{\partial t} 
+ \left(Ax-BR^{-1}B'\Pi x - BR^{-1}B'
\beta_j\right)'\frac{\partial g_j}{\partial x}\\
&+\frac{1}{2} \text{Tr} \left( \sigma' 
\frac{\partial^2 g_j}{\partial x^2} \sigma \right)=0.
\end{align*}}{
\[
\frac{\partial g_j}{\partial t} 
+ \left(Ax-BR^{-1}B'\Pi x - BR^{-1}B'
\beta_j\right)'\frac{\partial g_j}{\partial x}
+\frac{1}{2} \text{Tr} \left( \sigma' 
\frac{\partial^2 g_j}{\partial x^2} \sigma \right)=0.
\]
}
Hence, 
\begin{equation} \label{parabolic}
\frac{\partial \psi }{\partial t} 
+ 
x'A'\frac{\partial \psi}{\partial x}+\frac{1}{2} \text{Tr} \left( \sigma' 
\frac{\partial^2 \psi}{\partial x^2} \sigma \right)
-\eta\|x-\bar{x}\|^2_Q
\psi=0.
\end{equation}
By multiplying the right and left-hand sides of \eqref{parabolic} by $\frac{1}{\eta}\exp(\eta V(t,x))$, $V(t,x)$ satisfies (\ref{general_HJB}). 
The uniqueness of the solution follows from the uniqueness of
solutions to the uniform parabolic PDE (\ref{parabolic}) \cite[Theorem 7.6]{karatzas2012brownian}.

\subsection*{Proof of Theorem \ref{thm-opt}} \label{Proof5}

We have 
\iftoggle{jou}{
\begin{align*}
&u_*(t,x)=-R^{-1}B'\frac{\partial V}{\partial x}\\
&=\sum_{j=1}^l\frac{\psi_j(t,x)
g_j(t,x)}{\sum_{k=1}^l\psi_k(t,x)
g_k(t,x)}u^{(j)}(t,x)\\
&+\frac{1}{\eta\sum_{k=1}^l\psi_k(t,x)
g_k(t,x)}R^{-1}B'\sum_{j=1}^l
\psi_j(t,x)\frac{\partial g_j}{\partial x}(t,x).
\end{align*}}{
\begin{align*}
u_*(t,x)&=-R^{-1}B'\frac{\partial V}{\partial x}=\sum_{j=1}^l\frac{\psi_j(t,x)
g_j(t,x)}{\sum_{k=1}^l\psi_k(t,x)
g_k(t,x)}u^{(j)}(t,x)\\
&+\frac{1}{\eta\sum_{k=1}^l\psi_k(t,x)
g_k(t,x)}R^{-1}B'\sum_{j=1}^l
\psi_j(t,x)\frac{\partial g_j}{\partial x}(t,x).
\end{align*}
}
In the following we show that the second summand is zero.
By the change of variable $z=y-\alpha(T,t)x+
\int_t^T \alpha(T,\tau)BR^{-1}B'\beta_j(\tau)d\tau$ in \eqref{g1_g2 bis} and Leibniz integral rule, we have   
%See also perhaps Cortes et al. for possibly available results on these type of derivatives for Voronoi
\iftoggle{jou}{
\begin{align*}
&\frac{\partial g_j}{\partial x}(t,x)
=\frac{-\alpha(T,t)}{\sqrt{|2\pi\Sigma_t|}}
\int_{\partial \mathcal{W}_j-\alpha(T,t)x+
\int_t^T \alpha(T,\tau)BR^{-1}B'\beta_j(\tau)d\tau }\\ 
&\exp \left ( 
-\left\|z
\right\|^2_{\Sigma^{-1}_t} 
\right ) \vec{n}_j(z) ds(z)=\frac{-\alpha(T,t)}{\sqrt{|2\pi\Sigma_t|}}
\int_{\partial \mathcal{W}_j} \exp \Bigg ( \\ 
&-\left\|y-\alpha(T,t)x+
\int_t^T \alpha(T,\tau)BR^{-1}B'\beta_j(\tau)d\tau 
\right\|^2_{\Sigma^{-1}_t} 
\Bigg )\\
&\times \vec{n}_j(y) ds(y),
\end{align*}}{
\begin{align*}
&\frac{\partial g_j}{\partial x}(t,x)
=\frac{-\alpha(T,t)}{\sqrt{|2\pi\Sigma_t|}}
\int_{\partial \mathcal{W}_j-\alpha(T,t)x+
\int_t^T \alpha(T,\tau)BR^{-1}B'\beta_j(\tau)d\tau } \exp \left ( 
-\left\|z
\right\|^2_{\Sigma^{-1}_t} 
\right ) \vec{n}_j(z) ds(z)\\
&=\frac{-\alpha(T,t)}{\sqrt{|2\pi\Sigma_t|}}
\int_{\partial \mathcal{W}_j} \exp \left ( 
-\left\|y-\alpha(T,t)x+
\int_t^T \alpha(T,\tau)BR^{-1}B'\beta_j(\tau)d\tau 
\right\|^2_{\Sigma^{-1}_t} 
\right )\vec{n}_j(y) ds(y),
\end{align*}
}
where $\alpha$ is defined in \eqref{eq: definition of alpha} and $\vec{n}_j(y)$ is the unit normal component of $\partial \mathcal{W}_j$ and its translation $\partial \mathcal{W}_j-\alpha(T,t)x+
\int_t^T \alpha(T,\tau)BR^{-1}B'\beta_j(\tau)d\tau$.
By solving for $\beta_j$ and $\delta_j$ in (\ref{pi-beta-delta})
and replacing the solutions in the expressions
of the costs $V_j$ defined in \eqref{aux_cost} and in the 
derivatives $\frac{\partial g_j}{\partial x}$, one can show that under Assumption \ref{assumption_linear},
\iftoggle{jou}{
\begin{align}\label{cons_prob}
&\sum_{j=1}^l
\psi_j(t,x)\frac{\partial g_j}{\partial x}(t,x)
=K_1(t,x)
\sum_{j=1}^l\\
&\int_{\partial \mathcal{W}_j}
\exp\left(K_2(t,x,y) +\eta \|y-p_j\|_M^2
-\eta\|y\|_M^2
\right)\vec{n}_j(y) ds(y).\nonumber
\end{align}}{
\begin{align} \label{cons_prob}
&\sum_{j=1}^l
\psi_j(t,x)\frac{\partial g_j}{\partial x}(t,x)
=K_1(t,x)
\sum_{j=1}^l\int_{\partial \mathcal{W}_j}
\exp\left(K_2(t,x,y) +\eta \|y-p_j\|_M^2
-\eta\|y\|_M^2
\right)\vec{n}_j(y) ds(y).
\end{align}
}
where $K_1$ and $K_2$ are functions that do not depend on $p_j$,
$\forall j \in \{1,\dots,l\}$. Note that $\partial \mathcal{W}_j = \cup_{i=1}^{k_j} O_i$, where the disjoint
subsets (up to a subset of measure zero) $\{O_i\}_{i=1}^{k_j}$ 
are the common boundaries of $\mathcal{W}_j$
and the adjacent Voronoi cells. If 
$O_i$ is the common boundary of $\mathcal{W}_j$
and some adjacent Voronoi Cell $\mathcal{W}_k$, then
$\vec{n}_j(y)=-\vec{n}_k(y)$ for all $y \in O_i$. Moreover,
by the definition of the Voronoi cells,
$\|y-p_j\|_M=\|y-p_k\|_M$ for all $y \in 
O_i$. Therefore, the right-hand side of \eqref{cons_prob} is equal to zero.
Thus, the optimal control $u_*$ satisfies \eqref{optimal-crt-exp}. 

Next, we show that $u_*$ is an admissible Markov policy, i.e. $u_* \in \mathcal{L}$ defined in \eqref{set_l}. 
In view of (\ref{optimal-crt-exp}), the function $\frac{\partial u_*}{\partial x}$ is 
continuous on $[0,T)\times \mathbb{R}^n$. Therefore, the local Lipschitz
condition holds. Moreover, for 
all $(t,x) \in [0,T] \times \mathbb{R}^n$,
we have
\iftoggle{jou}{
\begin{align}  \label{linear_bound_ctr}
\|u_*(t,x)\| &\leq \sum_{j=1}^l\left\|u_*^{(j)}(t,x)\right\| \nonumber\\
&\leq 
\|R^{-1}B'\| \left(l\|\Pi\|_\infty 
\|x\| +
\sum_{j=1}^l\|\beta_j\|_\infty \right).
\end{align}}{
\begin{equation}  \label{linear_bound_ctr}
\|u_*(t,x)\| \leq \sum_{j=1}^l\left\|u_*^{(j)}(t,x)\right\| \leq 
\|R^{-1}B'\| \left(l\|\Pi\|_\infty 
\|x\| +
\sum_{j=1}^l\|\beta_j\|_\infty \right).
\end{equation}
}
Hence, the linear growth condition 
is satisfied and this proves the result.
As a result, sufficient conditions are satisfied for the 
SDE defined in (\ref{min-LQG}) and controlled by $u_*(t,x)$ to have
unique strong solution denoted $x_*(.)$ \cite[Section 5.2]{karatzas2012brownian}.

Finally, by the verification theorem \cite[Theorem 4.3.1]{fleming2006controlled}, we know that $u_*$ is  the unique  optimal control law of \eqref{min-LQG}
if it is the unique minimizer (up to a set of measure $0$) of the Hamiltonian $H(x,\frac{\partial V}{\partial x},u,t)= (Ax+Bu)'\frac{\partial V}{\partial x}+\|x-\bar{x}\|_Q^2+\|u\|_R^2$, and if the cost-to-go $V(t,x)$ has a polynomial growth in $x$ and satisfies \eqref{general_HJB}. 
For the first condition, we have for $\text{Lebesgue}\times \mathbb{P}$-a.e $(t,\xi) \in [0,T] \times \Omega$ ($\mathbb{P}$ is the probability measure defined at the beginning of Section \ref{section_model}),
\begin{align*}
u_*\left(t,x_*(t,\xi)\right)&=-R^{-1}B'\frac{\partial V}{\partial x}\left(t,x_*(t,\xi)\right)
\\&= \argmin\limits_{u \in 
\mathbb{R}^n} 
H\left(x_*(t,\xi),\frac{\partial V}{\partial x}(t,x_*(t,\xi)),u,t\right).
\end{align*}
In fact, the control law defined in \eqref{optimal_ctr} minimizes $H$ except on the set $\{T\} \times \Omega$, which has a $\text{Lebesgue}\times \mathbb{P}$ measure zero. Next,
in view of (\ref{linear_bound_ctr}), we have for all 
$(t,x) \in [0,T)\times \mathbb{R}^n$
\[
\|V(t,x)\| \leq \int_0^{\|x\|} \left\|\frac{\partial V}{\partial x}(t,y) \right\| dy \leq 
K_1(1+\|x\|^2),
\]
for some $K_1>0$. Moreover, 
$\|V(T,x)\| \leq K_2(1+\|x\|^2)$, for some $K_2>0$.
Hence,
for all $(t,x) \in [0,T]\times \mathbb{R}^n$,  
$
\|V(t,x)\| \leq K\left(1+\|x\|^2\right)$,
for some $K>0$.
Moreover, as established in Theorem 3, $V \in C^{1,2}\left([0,T)\times \mathbb{R}^n\right) \cap C([0,T] \times \mathbb{R}^n)$ satisfies the HJB equation
(\ref{general_HJB}). This proves the 
result.

\subsection{Proof of Lemma \ref{contr}} \label{Proof51}

To prove the result, it is sufficient to show that the expectation of the optimal
cost
$\mathbb{E}J_*(x(0)) \leq K+\frac{\sigma^2}{2}\log M$, for some $K>0$ independent of $M$. The result is then a direct consequence of Chebyshev's inequality 
\[
\mathbb{P}\left ( \min\limits_{1 \leq j \leq l}|
x_*(T)-p_j| > \epsilon\right)
\leq \frac{1}{\epsilon^2} \mathbb{E} \min\limits_{1 \leq j \leq l}|
x_*(T)-p_j|^2\leq \frac{2}{M\epsilon^2}
\mathbb{E}J_*(x(0)).\] To prove the
boundedness of the cost, we
start by the special 
case where $\bar{x}=0$ and $p_1=\dots=p_l=0$. The optimal cost is then
$J_*(x(0))=\frac{1}{2} \Pi(0) x^2(0)+
\frac{\sigma^2}{2}\int_0^T \Pi(\tau) d\tau$.
We now show that $\Pi(0)$ is uniformly
bounded with $M$ and that $\int_0^T \Pi(\tau)d\tau$
is of the order $\log M$ for large $M$.
To prove the uniform boundness of
$\Pi(0)$, we consider the LQR problem
\begin{equation} \label{LQR_cont}
\begin{split}
&\inf\limits_{v} 
J_M\left(y(0)=1,v(.)\right)=
\inf\limits_{v}
\int_0^T \left\{ \frac{Q}{2} y^2
+ \frac{R}{2} v^2 \right\} dt + \frac{M}{2}
 |y(T)|^2 \\
&\text{s.t. }
\frac{dy}{dt}= Ay + Bv, \qquad y(0)=1.
\end{split}
\end{equation}
The optimal cost is $J_M^*=\frac{1}{2}\Pi(0)$, where $\Pi$ is defined in (\ref{pi-beta-delta}). By controllability of $(A,B)$, one
can find a continuous control law $v_{10}$
that does not depend on $M$ and such that
the corresponding state $y_{10}$ is at time
$T$ at $0$. We have $\frac{1}{2}\Pi(0)=J_M^* \leq J_M (1,v_{10})$. The right hand 
side of the inequality is finite and does not depend on $M$. Hence, $\Pi(0)$ is uniformly bounded w.r.t. $M$. We now prove 
that $\int_0^T \Pi(\tau)d\tau$ is of the order $\log M$ for large $M$. 
We have
for all $M \geq 1$, $\frac{1}{2}\Pi(0) =
J_{M}^* \geq J_{M=1}^* := C>0$. The constant $C$ is independent of $M$, whenever $M >1$. Moreover, for large, $M$
$\min\limits_{t \in [0,T]} \Pi(t)=\Pi (0)
>C$. By dividing by $\Pi(t)$ on both sides of the Riccati equation in (\ref{pi-beta-delta}) and integrating on $[0,T]$ the 
right and left hand sides, we get
\[
\log M - \log \Pi(0) = \frac{b^2}{r} 
\int_0^T \Pi (\tau) d\tau -2aT -q \int_0^T
\frac{1}{\Pi(\tau)} d \tau. 
\]
By the boundedness of $\Pi(0)$ and $\left |\int_0^T
\frac{1}{\Pi(\tau)} d \tau \right| \leq \frac{1}{CT} $, we have $\int_0^T \Pi (\tau) d\tau / \log M$ converges to $1$ as
$M$ goes to infinity. Having shown the result for the special case, the case where
$\bar{x} \neq 0$ and $p_1=\dots=p_l=p$ can be proved by making the change of variables
$\tilde{x}=x-p$ and $\tilde{u}=u+\frac{A}{B}p$ and noting the uniform boundedness
of $\bar{x}$ and that
\[
\mathbb{E}J_*(x(0)) \leq \mathbb{E} \inf\limits_{\tilde{u}} 
\mathbb{E}\left (\int_0^T \{Q\tilde{x}^2+
R\tilde{u}^2\}dt + \frac{M}{2} \tilde{x}^2
(T)| x(0) 
 \right)+\int_0^T Q(\bar{x}-p)^2 dt+\frac{RA^2T}{B^2}p^2.\]
Finally, we conclude the general case  by the following inequality
\[
\mathbb{E}J_*(x(0)) \leq \mathbb{E} \inf\limits_{u} 
\mathbb{E}\left (\int_0^T \{\frac{Q}{2}(x-\bar{x})^2+
\frac{R}{2}u^2\}dt + \frac{M}{2} (x(T)
-p_1)^2 | x(0) 
 \right).
\]

\section{} \label{ProofB}  

This appendix includes the 
proofs of lemmas and theorems 
related to the existence of a  solution of the mean field 
equations.

\subsection*{Proof of Lemma \ref{lem-equ}} \label{Proof7}

First, we provide in the following lemma a stochastic maximum   %\ref{stochastic_max}
principle \cite{peng1990general} for the ``min-LQG'' optimal control problem. 
Because of the non-smooth final cost, this result is derived using the relationship
between dynamic programming and the stochastic maximum principle rather than 
the variational method used in \cite{peng1990general}.
\begin{lem} \label{stochastic_max}
The processes
$\left(q^s(t), \frac{\partial^2 V^s}{\partial x^2}(t,x_*^s(t)\right)$, 
$1\leq s \leq k$, with $q^s(t)=\frac{\partial V^s}{\partial x}(t,x_*^s(t))$, satisfy
the following backward linear SDE:
\iftoggle{jou}{
\begin{align} \label{adjoint_process}
-dq^s(t)&= \left( (A^s)^' q^s(t) +Q^s (x_*^s(t)-\bar{x}(t))
\right)dt\nonumber\\
&-  \frac{\partial^2 V^s}{\partial x^2}
\left((t,x_*^s(t)\right)\sigma^s dw^s(t), 
\end{align}}{
\begin{equation} \label{adjoint_process}
-dq^s(t)= \left( (A^s)^' q^s(t) +Q^s (x_*^s(t)-\bar{x}(t))
\right)dt
-  \frac{\partial^2 V^s}{\partial x^2}
\left((t,x_*^s(t)\right)\sigma^s dw^s(t), 
\end{equation}
}
with $q^s(T)=M^s\left(x_*^s(T)-
\sum_{j=1}^l 1_{\mathcal{W}_j}(x_*^s(T))p_j \right)$.
%Moreover, $\bar{n}(t)=\mathbb{E}n(t)$ satisfies on $[0,T)$ the 
%following linear backward ODE:
%\begin{equation}  \label{backward_adjoint_process}
%\frac{d}{dt}\bar{n}(t) = -a\bar{n}(t)+q
%(\bar{x}_*(t)-\bar{x}(t)),
%\end{equation}
%with $\bar{n}(T):=\lim\limits_{t \rightarrow T}\bar{n}(t)=M(\bar{x}_*(T)-p_\lambda)$, 
%where $\lambda=\mathbb{P}(x_*(T)<c)$ and 
%$p_\lambda= \lambda p_1 + (1-\lambda)p_2$.
\end{lem}
\begin{IEEEproof}
The function $\frac{\partial V^s}{\partial x}(t,x)$ is smooth
on $[0,T)\times \mathbb{R}^n$. By applying It\^o's formula
\cite[Section 3.3.A]{karatzas2012brownian} to 
$\frac{\partial V^s}{\partial x}\left(t,x_*^s(t)\right)$, 
and by noting that $V^s$ satisfies the HJB equation
(\ref{general_HJB}), we have
\iftoggle{jou}{
\begin{align*}
-dq^s(t)&= \left( (A^s)' q^s(t) +Q^s (x_*^s(t)-\bar{x}(t))
\right)dt\\
&- \frac{\partial^2 V^s}{\partial x^2}
\left((t,x_*^s(t)\right)\sigma^s dw^s(t),
\end{align*}}{
\[
-dq^s(t)= \left( (A^s)' q^s(t) +Q^s (x_*^s(t)-\bar{x}(t))
\right)dt
- \frac{\partial^2 V^s}{\partial x^2}
\left((t,x_*^s(t)\right)\sigma^s dw^s(t),
\]}
with 
$q^s(0)=\frac{\partial V^s}{\partial x}\left(0,x_*^s(0)\right)$. It 
remains to show that $\mathbb{P}$-a.s
\begin{equation} \label{limit_proof}
\lim\limits_{t \rightarrow T}\frac{\partial V^s}{\partial x}\left(t,x_*^s(t)\right)=M^s\left(x_*^s(T)-
\sum_{j=1}^l 1_{\mathcal{W}_j}(x_*^s(T))p_j \right).
\end{equation}
By Theorem \ref{thm-opt}, we have
on $[0,T)\times \mathbb{R}^n$
\begin{equation*} 
\frac{\partial V^s}{\partial x}(t,x)=\sum_{j=1}^l\frac{\exp\left(- 
\eta^s V_j^s(t,x) \right)
g_j^s(t,x)}{\sum_{k=1}^l\exp\left(- 
\eta^s V_k^s(t,x) \right)
g_k^s(t,x)}\frac{\partial V_j^s}{\partial x}(t,x).
\end{equation*}
Fix $j \in \{ 1, \ldots, l\}$. By Lemma \ref{lemma_g}, we have
on $\{ x_*^s(T) \in \text{Int}(\mathcal{W}_j) \}$,
$\lim\limits_{t\rightarrow T} g_j^s(t,x_*^s(t))=1$
and $\lim\limits_{t\rightarrow T} g_k^s(t,x_*^s(t))=0$, for all $k \neq j$. Hence, on $\{ x_*^s(T) \in \text{Int}(\mathcal{W}_j) \}$, we have $\lim\limits_{t\rightarrow T} \frac{\partial V^s}{\partial x}(t,x_*^s(t))=\lim\limits_{t\rightarrow T} \frac{\partial V_j^s}{\partial x}(t,x_*^s(t))=M^s(x_*^s(T)-p_j)$. 
But, $x_*^s$ is the solution of an
SDE with non degenerate noise. Therefore,
$\mathbb{P}\left( x_*^s(T) \in \partial{\mathcal{W}}_j \right)=0$.
Hence, (\ref{limit_proof}) holds. 
\end{IEEEproof}

\begin{rem}
The backward SDE (\ref{adjoint_process}) is the adjoint equation \cite{peng1990general} 
%related to 
for the min-LQG optimal control problem.
\end{rem}

We now prove Lemma \ref{lem-equ}.
By taking the expectations on the 
right and the left hand sides of (\ref{adjoint_process}) and the SDE in (\ref{gen_SDE}), 
and in view of $\sum_{s=1}^k \rsa{\alpha}_s \bar{x}^s(t)=\bar{x}$, we get the necessary
condition. To prove the sufficient 
condition, we consider $(\bar{X},\bar{x},\bar{q})$ 
satisfying (\ref{MF-1})-(\ref{MF-4}). We define 
$(\hat{x}^s,\hat{q}^s)=(\mathbb{E}x_*^s,\mathbb{E}q^s)$, 
where $(x_*^s,q^s)$ are the $s$-type generic agent's optimal state 
and co-state when tracking $\bar{x}$. 
We define $e=(\hat{x}^1,\dots,\hat{x}^k)-\bar{X}$ and 
$\bar{q}_e=(\hat{q}^1,\dots,\hat{q}^k)-\bar{q}$.
% \jlnComment{Notation is now pretty 
% confusing. We have
% $\bar x^s$, $\bar x^s_*$, $x^s_*$.}
By taking expectations on the right and the left hand sides 
of (\ref{adjoint_process}) and the generic agent's dynamics, 
we obtain that  
\begin{equation} \label{equ_lem9} 
\begin{split}
\frac{d}{dt}e(t)&=Ae(t)-BR^{-1}B'
\bar{q}_e(t), \qquad e(0)=0\\
\frac{d}{dt} \bar{q}_e(t)&=-A' \bar{q}_e(t)
+QLe(t),  \qquad
\bar{q}_e(T)=Me(T).
\end{split}
\end{equation}
Under Assumption \ref{assum_riccati}, we define $q'_e(t)=\pi(t) e(t)$, where
$\pi(t)$ is the unique solution of \eqref{aux_riccati}. We have
$\frac{d(\bar{q}_e-q'_e)}{dt}=-(A'-\pi(t)BR^{-1}B')(\bar{q}_e-q'_e)$, with 
$(\bar{q}_e(T)-q'_e(T))=0$. Hence, $\bar{q}_e(t)=\pi(t)e(t)$. 
By replacing $\bar{q}_e(t)=\pi(t)e(t)$ in the
forward equation in \eqref{equ_lem9}, we obtain that $e=0$. 
Hence, $\bar{x}$ satisfies (\ref{gen_SDE}).
%\jlnComment{just say: $\bar x$ satisfies (17)?}

\subsection*{Proof of Theorem \ref{lemma:existance11}} \label{Proof8}

Let $\bar{x}$ be a path satisfying (\ref{gen_SDE}). Then, by Lemma \ref{lem-equ}, 
$\bar{x}$ satisfies (\ref{MF-1})-(\ref{MF-4}). Under Assumption \ref{assum_riccati}, using arguments similar to those used in Lemma \ref{lem-equ} we obtain that
(\ref{MF-1}) and (\ref{MF-2})
has a unique solution $(\bar{X},\bar{q})$. Moreover,
$\bar{q}=\pi\bar{X}+\gamma$, where
$\pi$ is the unique solution of 
\eqref{aux_riccati}, and $\gamma$
is the unique solution of 
$\dot{\gamma}=-(A-BR^{-1}B'\pi)'\gamma$ with $\gamma(T)=-M\Lambda \otimes I_n p$. By replacing, $\bar{q}=\pi\bar{X}+\gamma$ in (\ref{MF-1}), we get that
$\bar{x}$ is of the form (\ref{eq: x-lambda def}). 
Next, by implementing this new form of $\bar{x}$ in the expression of (\ref{MF-4}) and 
by noting that $\Lambda$ satisfies (\ref{MF-3}), $\Lambda$ is a fixed point of
$F$. Conversely, we consider $\Lambda$ to be a fixed
point of $F$, $\bar{X}=\left(R_1(t,0)\bar{X}(0)+R_2(t)\Lambda \otimes I_n p\right)$ 
and $\bar{x}=P_1\bar{X}$. We define  
$\bar{q}(t)= -(BR^{-1}B')^{-1}(\frac{d}{dt}\bar{X}(t)-A \bar{X}(t))$. $(\bar{X},\bar{q})$ satisfies (\ref{MF-1})-(\ref{MF-2}).
We have $\Lambda_{sj}=F(\Lambda)_{sj}=\mathbb{P}(x_*^{s,\Lambda}(T) \in \mathcal{W}_j)$,
where $x_*^{s,\Lambda}$ is defined in (\ref{finite-operator}). 
But $\bar{x}$ is of the form (\ref{form_fixed}), hence $x_*^\lambda$
is the unique strong solution of (\ref{MF-4}).
Therefore, $\bar{x}$ satisfies 
(\ref{MF-1})-(\ref{MF-4}), and by Lemma \ref{lem-equ}, it satisfies (\ref{gen_SDE}). This proves the first point.

Next, to show the existence of a fixed point of $F$, it is sufficient to show that $F$ is continuous, in which case 
Brouwer's fixed point theorem \cite[Section V.9]{FunAnaCon} ensures the existence of a fixed point. 
Equation (\ref{finite-operator}) is a stochastic differential equation depending 
on the parameter $\Lambda$. By \cite[Theorem 1]{skorokhod1981stochastic}, the 
joint distribution of $X_*^\Lambda$ and the Brownian motion $W$
is weakly continuous in $\Lambda$. Consider a sequence of stochastic matrices $\{\Lambda_n\}_{n\geq 0}$ converging to the stochastic matrix $\Lambda$. The distribution of $X_*^{\Lambda_n}(T)$ converges 
weakly to the distribution of $X_*^\lambda(T)$ 
%\rmaComment{The stochastic processes converge weakly/vaguely to each other, but do the distributions converge weakly?}. 
Moreover, $X_*^\lambda$ is the solution of a non-degenerate SDE. Hence, $\mathcal{W}_j$,
$j=1,\dots,l$, is a continuity set of the 
distribution of $X_*^\lambda$. Therefore,
$\lim\limits_n F(\Lambda_n)_{sj}=\lim\limits_n \mathbb{P}(x_*^{s,\Lambda_n}(T)\in \mathcal{W}_j)=
\mathbb{P}(x_*^{s,\Lambda}(T)\in \mathcal{W}_j)=F(\Lambda)_{sj}$, 
and so $F$ is continuous.
 Finally, using \rma{arguments similar to} those used in \cite[Lemma 9]{DBLP:journals/corr/SalhabMN15}, one can show the third point.

\ifCLASSOPTIONcaptionsoff
  \newpage
\fi

\bibliographystyle{IEEEtran}
\bibliography{IEEEabrv,mfg} 

\begin{thebibliography}{10}
\providecommand{\url}[1]{#1}
\csname url@samestyle\endcsname
\providecommand{\newblock}{\relax}
\providecommand{\bibinfo}[2]{#2}
\providecommand{\BIBentrySTDinterwordspacing}{\spaceskip=0pt\relax}
\providecommand{\BIBentryALTinterwordstretchfactor}{4}
\providecommand{\BIBentryALTinterwordspacing}{\spaceskip=\fontdimen2\font plus
\BIBentryALTinterwordstretchfactor\fontdimen3\font minus
  \fontdimen4\font\relax}
\providecommand{\BIBforeignlanguage}[2]{{%
\expandafter\ifx\csname l@#1\endcsname\relax
\typeout{** WARNING: IEEEtran.bst: No hyphenation pattern has been}%
\typeout{** loaded for the language `#1'. Using the pattern for}%
\typeout{** the default language instead.}%
\else
\language=\csname l@#1\endcsname
\fi
#2}}
\providecommand{\BIBdecl}{\relax}
\BIBdecl

\bibitem{Kappelman2005}
F.~Koppelman and V.~Sathi, ``Incorporating variance and covariance
  heterogeneity in the generalized nested logit model: an application to
  modeling long distance travel choice behavior,'' \emph{Transportation
  Research}, vol.~39, pp. 825--853, 2005.

\bibitem{Bhat2004}
C.~Bhat and J.~Guo, ``A mixed spatially correlated logit model: formulation and
  application to residential choice modeling,'' \emph{Transportation Research},
  vol.~38, pp. 147--168, 2004.

\bibitem{nakajima2007measuring}
R.~Nakajima, ``Measuring peer effects on youth smoking behaviour,'' \emph{The
  Review of Economic Studies}, vol.~74, no.~3, pp. 897--935, 2007.

\bibitem{vail2003multi}
D.~Vail and M.~Veloso, ``Multi-robot dynamic role assignment and coordination
  through shared potential fields,'' \emph{Multi-robot systems}, pp. 87--98,
  2003.

\bibitem{halasz2007dynamic}
A.~Hal{\'a}sz, M.~A. Hsieh, S.~Berman, and V.~Kumar, ``Dynamic redistribution
  of a swarm of robots among multiple sites,'' in \emph{2007 IEEE/RSJ
  International Conference on Intelligent Robots and Systems}.\hskip 1em plus
  0.5em minus 0.4em\relax IEEE, 2007, pp. 2320--2325.

\bibitem{hsieh2008biologically}
M.~A. Hsieh, {\'A}.~Hal{\'a}sz, S.~Berman, and V.~Kumar, ``Biologically
  inspired redistribution of a swarm of robots among multiple sites,''
  \emph{Swarm Intelligence}, vol.~2, no. 2-4, pp. 121--141, 2008.

\bibitem{Merill99_vote}
S.~Merrill and B.~Grofman, \emph{A Unified Theory of Voting: Directional and
  Proximity Spatial Models}.\hskip 1em plus 0.5em minus 0.4em\relax Cambridge
  University Press, 1999.

\bibitem{markus1979dynamic}
G.~B. Markus and P.~E. Converse, ``A dynamic simultaneous equation model of
  electoral choice,'' \emph{American Political Science Review}, vol.~73,
  no.~04, pp. 1055--1070, 1979.

\bibitem{mcfadden1973conditional}
D.~McFadden, ``Conditional logit analysis of qualitative choice behavior,'' in
  \emph{Frontiers in Econometrics}, P.~Zarembka, Ed.\hskip 1em plus 0.5em minus
  0.4em\relax New York: New York: Academic Press, 1974, ch.~4, pp. 105--142.

\bibitem{rust1994structural}
J.~Rust, ``Structural estimation of markov decision processes,'' \emph{Handbook
  of econometrics}, vol.~4, pp. 3081--3143, 1994.

\bibitem{Brock2001}
W.~Brock and S.~Durlauf, ``Discrete choice with social interactions,''
  \emph{Review of Economic Studies}, pp. 147--168, 2001.

\bibitem{Rab2014}
R.~Salhab, R.~P. Malham\'e, and J.~Le~Ny, ``Consensus and disagreement in
  collective homing problems: A mean field games formulation,'' in
  \emph{Proceedings of the 53rd IEEE Conference on Decision and Control}, Dec
  2014, pp. 916--921.

\bibitem{Rab2015}
------, ``A dynamic game model of collective choice in multi-agent systems,''
  in \emph{Proceedings of the 54th IEEE Conference on Decision and Control},
  Dec 2015, pp. 4444--4449.

\bibitem{DBLP:journals/corr/SalhabMN15}
R.~Salhab, R.~P. Malham{\'e}, and J.~Le~Ny, ``A dynamic game model of
  collective choice in multi-agent systems,'' \emph{IEEE Transactions on
  Automatic Control}, 2017, {I}n Press, available at
  http://ieeexplore.ieee.org/document/7970149/.

\bibitem{Huang03_wirelessPower}
M.~Huang, P.~E. Caines, and R.~P. Malham\'e, ``Individual and mass behaviour in
  large population stochastic wireless power control problems: centralized and
  {N}ash equilibrium solutions,'' in \emph{Proceedings of the 42nd IEEE
  Conference on Decision and Control}, 2003, pp. 98--103.

\bibitem{Huang07_Large}
------, ``Large-population cost-coupled {LQG} problems with nonuniform agents:
  Individual-mass behavior and decentralized epsilon-{N}ash equilibria,''
  \emph{IEEE Transactions on Automatic Control}, vol.~52, no.~9, pp.
  1560--1571, 2007.

\bibitem{huang2006large}
M.~Huang, R.~P. Malham\'e, and P.~E. Caines, ``Large population stochastic
  dynamic games: closed-loop {McKean-Vlasov} systems and the {Nash} certainty
  equivalence principle,'' \emph{Communications in Information \& Systems},
  vol.~6, no.~3, pp. 221--252, 2006.

\bibitem{lasry2006jeux}
J.~M. Lasry and P.~L. Lions, ``Jeux {\`a} champ moyen. {I}--le cas
  stationnaire,'' \emph{Comptes Rendus Math{\'e}matique}, vol. 343, no.~9, pp.
  619--625, 2006.

\bibitem{lasry2006jeux2}
------, ``Jeux {\`a} champ moyen. {II}--horizon fini et contr{\^o}le optimal,''
  \emph{Comptes Rendus Math{\'e}matique}, vol. 343, no.~10, pp. 679--684, 2006.

\bibitem{Lasry07_MFG}
------, ``Mean field games,'' \emph{Japanese Journal of Mathematics}, vol.~2,
  pp. 229--260, 2007.

\bibitem{fagnani2008randomized}
F.~Fagnani and S.~Zampieri, ``Randomized consensus algorithms over large scale
  networks,'' \emph{IEEE Journal on Selected Areas in Communications}, vol.~26,
  no.~4, pp. 634--649, 2008.

\bibitem{karatzas2012brownian}
I.~Karatzas and S.~Shreve, \emph{Brownian motion and stochastic
  calculus}.\hskip 1em plus 0.5em minus 0.4em\relax Springer Science \&
  Business Media, 2012, vol. 113.

\bibitem{salhab2016dynamic}
R.~Salhab, R.~P. Malham{\'e}, and J.~Le~Ny, ``A dynamic game model of
  collective choice: Stochastic dynamics and closed loop solutions,''
  \emph{arXiv preprint arXiv:1604.08136}, 2016.

\bibitem{huang2012social}
M.~Huang, P.~E. Caines, and R.~P. Malham\'e, ``Social optima in mean field
  {LQG} control: centralized and decentralized strategies,'' \emph{IEEE
  Transactions on Automatic Control}, vol.~57, no.~7, pp. 1736--1751, 2012.

\bibitem{fleming2006controlled}
W.~H. Fleming and H.~M. Soner, \emph{Controlled Markov processes and viscosity
  solutions}.\hskip 1em plus 0.5em minus 0.4em\relax Springer Science \&
  Business Media, 2006, vol.~25.

\bibitem{evans1998partial}
L.~Evans, \emph{Partial Differential Equations}, ser. Graduate studies in
  mathematics.\hskip 1em plus 0.5em minus 0.4em\relax American Mathematical
  Society, 1998.

\bibitem{yong1999stochastic}
J.~Yong and X.~Y. Zhou, \emph{Stochastic controls: Hamiltonian systems and HJB
  equations}.\hskip 1em plus 0.5em minus 0.4em\relax Springer Science \&
  Business Media, 1999, vol.~43.

\bibitem{Gomes:Book16:regulatiryMFG}
D.~Gomes, E.~Pimentel, and V.~Voskanyan, \emph{Regularity Theory for Mean-Field
  Game Systems}, ser. SpringerBriefs in Mathematics.\hskip 1em plus 0.5em minus
  0.4em\relax Springer, 2016.

\bibitem{liggett2012interacting}
T.~Liggett, \emph{Interacting particle systems}.\hskip 1em plus 0.5em minus
  0.4em\relax Springer Science \& Business Media, 2012, vol. 276.

\bibitem{perko2013differential}
L.~Perko, \emph{Differential equations and dynamical systems}.\hskip 1em plus
  0.5em minus 0.4em\relax Springer Science \& Business Media, 2013, vol.~7.

\bibitem{anderson2007optimal}
B.~D. Anderson and J.~B. Moore, \emph{Optimal control: linear quadratic
  methods}.\hskip 1em plus 0.5em minus 0.4em\relax Dover Publications, 2007.

\bibitem{freiling2002survey}
G.~Freiling, ``A survey of nonsymmetric {R}iccati equations,'' \emph{Linear
  algebra and its applications}, vol. 351, pp. 243--270, 2002.

\bibitem{carmona2013probabilistic}
R.~Carmona and F.~Delarue, ``Probabilistic analysis of mean-field games,''
  \emph{SIAM Journal on Control and Optimization}, vol.~51, no.~4, pp.
  2705--2734, 2013.

\bibitem{pichler2013numerical}
L.~Pichler, A.~Masud, and L.~Bergman, ``Numerical solution of the
  {F}okker--{P}lanck equation by finite difference and finite element methods
  -— {A} comparative study,'' in \emph{Computational Methods in Stochastic
  Dynamics}.\hskip 1em plus 0.5em minus 0.4em\relax Springer, 2013, pp. 69--85.

\bibitem{durrett2010probability}
R.~Durrett, \emph{Probability: theory and examples}.\hskip 1em plus 0.5em minus
  0.4em\relax Cambridge university press, 2010.

\bibitem{peng1990general}
S.~Peng, ``A general stochastic maximum principle for optimal control
  problems,'' \emph{SIAM Journal on control and optimization}, vol.~28, no.~4,
  pp. 966--979, 1990.

\bibitem{FunAnaCon}
J.~B. Conway, \emph{A Course in Functional Analysis}, ser. Graduate Texts in
  Mathematics.\hskip 1em plus 0.5em minus 0.4em\relax Springer-Verlag, 1985.

\bibitem{skorokhod1981stochastic}
A.~V. Skorokhod, ``Stochastic differential equations depending on a
  parameter,'' \emph{Theory of Probability \& Its Applications}, vol.~25,
  no.~4, pp. 659--666, 1981.

\end{thebibliography}

\begin{IEEEbiography}[{\includegraphics[width=1.5in,height=1.25in,clip,keepaspectratio]{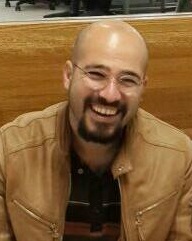}}]{Rabih Salhab} is currently pursuing the Ph.D. degree in Electrical Engineering in the Department of Electrical Engineering, Ecole Polytechnique de Montreal, Canada. He received the B.S. degree in Electrical Engineering from Ecole Sup\'erieure d'Ing\'enieurs de Beyrouth (E.S.I.B), Lebanon, in 2008. From 2008 to 2013 he was an 
Electrical Engineer with Dar al Handasah Shair and Partners, Lebanon. His current research interests are in stochastic control, game theory and mean field games theory and its applications.  
\end{IEEEbiography}

\begin{IEEEbiography}[{\includegraphics[width=1in,height=1.25in,clip,keepaspectratio]{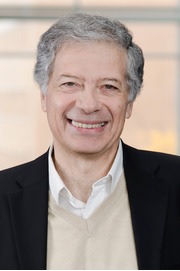}}]{Roland Malham\'e} received the Bachelor’s, Master’s and Ph.D. degrees in Electrical Engineering from the American University of Beirut, the University of Houston, and the Georgia Institute of Technology in 1976, 1978 and 1983 respectively.

After single year stays at University of Quebec , and CAE Electronics Ltd (Montreal), he joined in 1985 \'Ecole Polytechnique de Montr\'eal, where he is Professor of Electrical Engineering. In 1994, 2004, and 2012 he was on sabbatical leave respectively with LSS CNRS (France),  \'Ecole Centrale de Paris, and University of Rome Tor Vergata.
His interest in statistical mechanics inspired approaches to the analysis and control of large scale systems has led him to contributions in the area of aggregate electric load modeling, and to the early developments of the theory of mean field games. His current research interests are in collective decentralized decision making schemes, and the development of mean field based control algorithms in the area of smart grids. From june 2005 to june 2011, he headed GERAD, the Group for Research on Decision Analysis. He is an Associate Editor of International Transactions on Operations Research.
\end{IEEEbiography}

\begin{IEEEbiography}[{\includegraphics[width=1in,height=1.25in,clip,keepaspectratio]{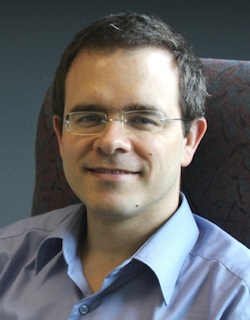}}]{Jerome Le Ny} (S'05-M'09-SM'16) received 
the Engineering Degree from the \'Ecole Polytechnique, France, in 2001, 
the M.Sc. degree in Electrical Engineering from the University of Michigan, Ann Arbor, in 2003, 
and the Ph.D. degree in Aeronautics and Astronautics from the Massachusetts Institute of Technology, 
Cambridge, in 2008.
He is currently an Associate Professor with the Department of Electrical Engineering, 
Polytechnique Montreal, Canada. 
He is a member of GERAD, a multi-university research center on decision analysis.  
From 2008 to 2012 he was a Postdoctoral Researcher with the GRASP Laboratory 
at the University of Pennsylvania. 
His research interests include robust and stochastic control, 
mean-field control, 
privacy and security in large-scale networked control systems, 
dynamic resource allocation and active perception,
with applications to autonomous multi-robot systems and intelligent infrastructures.
\end{IEEEbiography}

\end{document}